\documentclass[useAMS,usenatbib,letterpaper]{mn2e}
\usepackage{aas_macros,graphicx,times,multirow,amsmath,amssymb,color,longtable,pdflscape}
\title[The CATS@BAR project]{The \cxo\ ACIS Timing Survey Project: glimpsing a sample of faint X-ray pulsators}
\author[G.~L.~Israel et al.] {G.~L.~Israel,$^{1}$\thanks{E-mail: gianluca@oa-roma.inaf.it} P.~Esposito,$^{2,3}$ 
G.~A. Rodr\'iguez Castillo$ ^1$ and L.~Sidoli$^3$
\smallskip\\
$^1$INAF--Osservatorio Astronomico di Roma, via Frascati 33, I-00040 Monteporzio Catone, Italy\\
$^2${ Anton Pannekoek Institute for Astronomy, University of Amsterdam, Postbus 94249, NL-1090-GE Amsterdam, The Netherlands}\\
$^3$INAF--Istituto di Astrofisica Spaziale e Fisica Cosmica - Milano, via E. Bassini 15, I-20133 Milano, Italy
}
\date{Accepted 2016 July 27. Received 2016 July 23; in original form 2016 April 13} \pagerange{\pageref{firstpage}--\pageref{lastpage}} \pubyear{2016}

\def\LaTeX{L\kern-.36em\raise.3ex\hbox{a}\kern-.15em
    T\kern-.1667em\lower.7ex\hbox{E}\kern-.125emX}
\def\xmm {\emph{XMM--Newton}}
\def\cxo {\emph{Chandra}}
\def\swift {\emph{Swift}}

\def\igr {\emph{INTEGRAL}}

\def\rst {\emph{ROSAT}}
\def\asca {\emph{ASCA}}

\def\flux {\mbox{erg cm$^{-2}$ s$^{-1}$}}
\def\lum {\mbox{erg s$^{-1}$}}

\def\cbar {CATS\,@\,BAR}
\def\rc {\rm}
\def\rcb {\rm}
\def\core {\rm}
\begin{document}

\label{firstpage}
\maketitle
\begin{abstract}
We report on the discovery of  {\rc 41} new pulsating sources in the data of the \cxo\ Advanced CCD Imaging Spectrometer, which is sensitive to X-ray photons in the 0.3--10\,keV band. The archival data of the first 15 years of \cxo\ observations were retrieved and analysed by means of fast Fourier transforms, employing a peak-detection algorithm able to screen candidate signals in an automatic fashion. We carried out the search for new X-ray pulsators in light curves with more than 50 photons, for a total of about 190\,000 lightcurves out of about 430\,000 extracted. With these numbers, the ChAndra Timing Survey at Brera And Roma astronomical observatories (\cbar)---as we called the project---represents the largest ever systematic search for coherent signals in the classic X-ray band. More than 50 per cent of the signals were confirmed by further \cxo\ (for those sources with two or more pointings), \xmm\ or \rst\ data. The period distribution of the new X-ray pulsators above $\sim$2\,000\,s resembles that of cataclysmic variables, while there is a paucity of sources with shorter period and low fluxes. Since there is not an obvious bias against these detections, a possible interpretation is in terms of a magnetic gating mechanism in accreting neutron stars. Finally, we note that \cbar\ is a living project and the detection algorithm will continue to be routinely applied to the new \cxo\ data as they become public. Based on the results obtained so far, we expect to discover about three new pulsators every year.  
\end{abstract}


\section{Introduction}\label{intro}
The detection and characterization of X-ray periodic signals play a role of paramount importance in the process of identifying new compact objects or new classes of them, and studying the mechanisms powering the observed emission. Even new emission mechanisms have been discovered in this way. The greatest part of the periodic signals arise from the rotation of a compact star or from the orbital motion in a binary system. {\rc  \core The main cases} in which the modulation is due to the compact star spin are: (a) accreting magnetic neutron stars (NSs) in X-ray binary systems; (b) spinning down isolated NSs, the emission of which may be powered by the dissipation of rotational, thermal, or even magnetic energy (as in the cases of classical radio pulsars, X-ray dim isolated NSs, and magnetars, respectively); (c) magnetic white dwarf (WD) systems, such as polars and intermediate polars (IPs). Orbital modulations of the X-ray flux are observed in many classes of NS and black hole (BH) X-ray binaries and cataclysmic variables (CVs). 

The discovery of a periodic modulation generally happens through the timing analysis of the X-ray light curves of the sources which lie in the field of view, FoV, of a detector. However, for most observations taken with imaging instruments, only the target of the pointing is analysed. 
With the high energy missions currently operational (such as \cxo\ or \xmm), which have wide energy ranges, high sensitivities and, often, high time resolution, the average number of serendipitous X-ray sources detected {\rc is of the order of few tens.  In a typical \cxo\ observation this number is $\sim$40\footnote{{\rc This number is simply the ratio between detected sources, 430\,000, and analysed datasets, 10\,700; see also end of this Section.}}.} { \core During observations targeting crowded fields and in which all the CCDs were operating (each CCD covers 8.3\arcmin x 8.3\arcmin), up to about 100 sources can be detected}. It is therefore evident that up to 99\% of the information  remain unexplored. Since the number of high quality time series stored in the present-day X-ray archives has now reached $\sim$ $10^6$, these data clearly hold a huge potential for new discoveries and make data mining an urgent task to achieve them.  

Various projects aimed at the search for X-ray pulsators were carried out in the past. During the 1990s, our team carried out systematic timing analyses of the then-booming number of bright serendipitous sources---about 50\,000 objects detected with \rst\ and \emph{EXOSAT} {\rc \citep{israel98}}. The effort was based on discrete fast Fourier transform (FFT) analysis and resulted, among other findings, in the detection of pulsations from sources which have become prototypes of new classes, such as the anomalous X-ray pulsars (4U\,0142+614; \citealt{israel94}) and the double-degenerate ultra-short period X-ray binaries hosting two WDs (HM\,Cnc; \citealt{Israel02}). {\rc{ \core  Remarkably, these sources are supposed to be powered by new and unusual mechanisms, such as the decay of the intense magnetic field (see \citealt{turolla15} for a recent review) in the case of magnetars, and owing to the unipolar inductor model for ultra-compact binaries \citep{dallosso06}.}} We also discovered an X-ray pulsator in a binary systems in a previously unobserved evolutionary phase, a post-common envelope stage: HD\,497985 (RX\,J0648.0--4418). It is a 13-s-period source which hosts either one of the most massive and fastest spinning WDs observed so far, or an unusual accreting NS spinning up at a low and extremely steady rate (\citealt{israel97}; \citealt{mte09,mereghetti16}).

More recently, several searches were conducted by other teams on \cxo\ and \xmm\ data for a relatively limited number of objects or sky regions. \citet{mbb03} searched for periodic signals among 285 sources detected with \cxo\ in the Galactic center region (about 500\,ks of exposure time) by means of a $Z_1^2$ (Rayleigh) test \citep{buccheri83}. They found eight sources showing pulsations, with periods from $\sim$300 to 17\,000~s. 
\citet{muno08} analysed almost 1\,000 \cxo\ and \xmm\ archival observations of the Galactic plane region ($|b|<5\degr$). All the detected sources were searched for coherent signals with a Rayleigh test for the \cxo\ data and a discrete FFTs for \xmm\ data. Four sources with periodic modulations in the 200--5\,000~s range were discovered and tentatively classified as CVs. \citet{laycock10} analysed two \cxo\ deep fields in the Small Magellanic Cloud (SMC), where almost 400 sources were detected and searched for pulsations. Two new accreting pulsars in high mass X-ray binaries (HMXBs) were discovered using the Lomb--Scargle method (LS; \citealt{scargle82}). \citet{hvdbg12} searched about 400 sources in 1-Ms \cxo\ data of the `Limiting Window' region. A LS analysis led to the discovery of ten new X-ray sources with  periods between 4\,700 and 12\,100~s; in three cases the signal was detected with good confidence ($>$99\%). In a sample of 4\,330 sources of the Second \xmm\ Serendipitous Source Catalogue (2XMMi-DR3; \citealt{watson09}) which were reliably detected and observed at least twice, \citet*{lin14} identified about 200 compact object candidates. By a FFT analysis, they found two sources which also show significant coherent pulsations (in the 5\,000--8\,000\,s range). 
A Random Forest machine learning algorithm was employed by \citet{farrell15} to classify 2\,876 variable \xmm\ sources in the Third \xmm\ Serendipitous Source Catalogue (3XMMi-DR4; \citealt{rosen15}). Two out of the about 100 unclassified objects showed in LS periodograms significant modulations at 400 and 18\,000~s, identified with a spin and an orbital period, respectively.
{\rc Finally, { \core it} is worth mentioning the results obtained by { \core \emph{Fermi}} in the systematic { \core `blind'} search for $\gamma$-ray pulsars, { \core although} in a different energy band (about 100\,MeV--100\,GeV), by means of FFT analysis carried out on photon time differences instead of event times \citep{atwood06}. This method { \core made it possible} to discover more than 50 new $\gamma$-ray pulsars\footnote{{\rc For an { \core updated list,} see https://confluence.slac.stanford.edu/display/ GLAMCOG/Public+List+of+LAT-Detected+Gamma-Ray+Pulsars}.} \citep{abdo09}.}

In this paper, we report on the results of our search for new X-ray pulsators in all the archival data of the \cxo\ Advanced CCD Imaging Spectrometer (ACIS; \citealt{garmire03}) Imaging (I) or Spectroscopic (S) array in imaging mode. We analysed about 10\,700 datasets, corresponding to the first $\sim$15 years of \cxo\ operations, and applied a FFT analysis coupled with an {\it ad hoc} signal detection algorithm to about 190\,000 of about 430\,000 light curves extracted (those with enough photons, $>$50, to allow a meaningful search). The ChAndra Timing Survey at Brera And Roma astronomical observatories (\cbar), as we called the project, is the largest ever systematic search for coherent signals in the classic X-ray band. 
Among the aims of \cbar\ there are: (i) to identify intrinsically faint X-ray pulsators at a flux level which has remained unexplored so far; (ii) to extend the validity of already known emission mechanisms towards lower X-ray fluxes; (iii) to look for new classes of compact objects and new astrophysical objects showing coherent signals.
We found  {\rc 41} new X-ray pulsators, with periods in the 8--64\,000\,s range. For a relatively large fraction of them, we rely upon more than one archival pointing allowing us to check for the reliability of the signal detection. A by-product of our systematic search analysis is a detailed map of the spurious (instrumental) coherent signals which affect the ACIS. 

\section{Methodology}\label{methodology}
\begin{figure*}
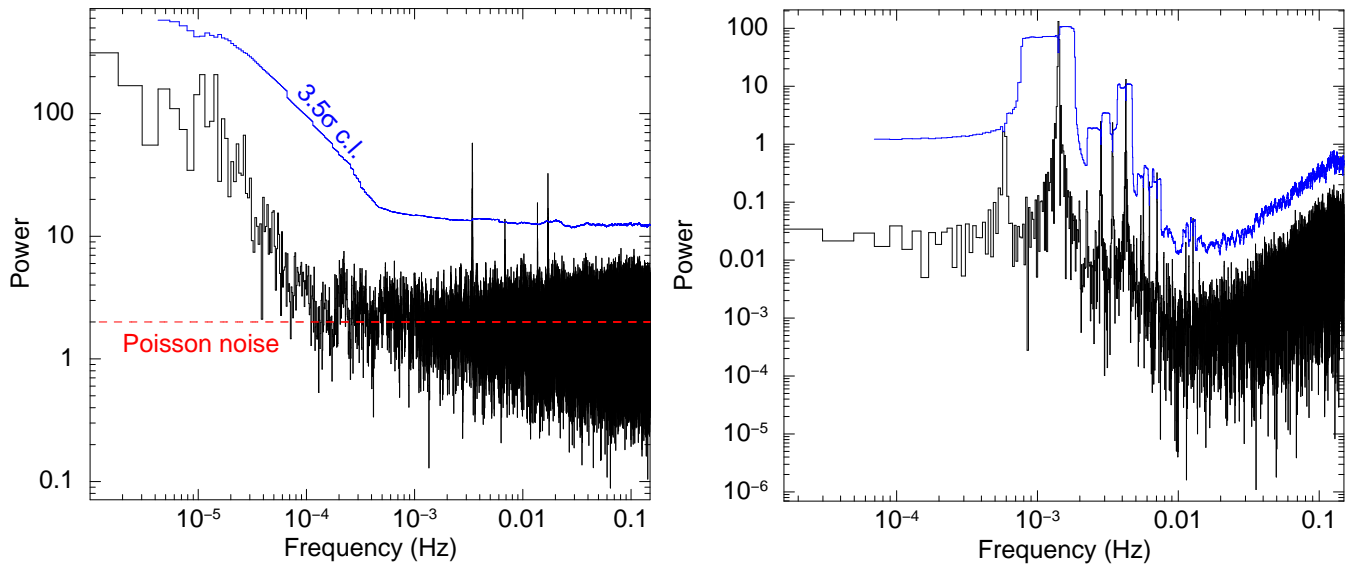

\centering
\resizebox{\hsize}{!}{\includegraphics[angle=-90]{dpsexample.ps}\hspace{.5cm}\includegraphics[angle=-90]{dpsdither.ps}}
\caption{\label{dps}  Left panel: Power spectrum density of the 292\,s-signal transient pulsar CXOU\,J005048.0--731817 in the SMC. The solid blue line indicates the 3.5$\sigma$ detection threshold normalised to the number of independent Fourier frequencies as described in Section\,\ref{methodology} {\rc and obtained by means of the algorithm described in \citet{israel96}}. The corresponding Poissonian noise level, according to the Leahy normalization \citep{leahy83}, is marked by the stepped red line. {\rc In this example a non Poissonian noise component, often referred as red-noise, is evident below 10$^{-4}$\,Hz when the power estimates  start deviating from the Poissonian noise level}. Right panel: As in the first panel, but for the time series of the effective area within the source extraction region as derived with the \textsc{ciao dither\_region} task.}
\end{figure*}

Fourier analysis is likely the single most important technique which is applied to astronomical time series to detect signals through the presence of peaks in the power spectra and characterise the noise variability through the study of continuum power spectrum components {\rcb (see \citealt{graham13} for a comparison among different signal search algorithms)}. A periodic signal is identified when the strength of the corresponding peak is so high to make it stand out in the power spectral density (PSD), and is validated (at a given confidence level) against the chance that the peak originates from the underlying white noise. This approach assumes that the power spectra are dominated by Poissonian counting statistics. However, this is often not the case, at least over a range of frequencies {\rc(see left panel of  Figure\,\ref{dps})}. Indeed, the presence of continuum non-Poissonian components in the power spectrum, resulting from intrinsic or instrumental variability in the source and/or the background, makes the detection of genuinely astrophysical signals a difficult statistical problem. Additionally, the number of serendipitous sources for which the timing analysis can be performed has became so large that it is virtually impossible to rely upon an inspection by eye of all the PSD produced in a large search to investigate every candidate signal. 

An algorithm to search for coherent signals in a large number of sources needs to include a reliable automatic procedure to perform a first screening of possible signals. 
{\rcb In our timing analysis we used the original (unbinned) event list extracted for each source in order to  avoid, among others, issues due to the starting time of the series (middle time of the bin versus start of the time bin), a problem often met in the astronomical timing software when binned light curves are used.
For the power spectrum density analysis} we adopted the recipe outlined in \citet{israel96}, which presents a statistical tool that can be employed for automatic screening of peaks {\rcb in PSD obtained by means of fast Fourier Transform (FFT)}. For each PSD, the algorithm infers a smoothed continuum model by maximizing the Kolmogorov--Smirnov probability that the power spectrum obtained by the ratio between the original one and the smoothed continuum is distributed as a $\chi^2$ with $2N$ degrees of freedom (dof), where $N$ is the number of averaged power spectra (in general, in our project $N=1$). In this way, it is possible to take into account statistically any non-Poissonian component and set a peak detection threshold which is frequency dependent. The left panel of Fig\,\ref{dps} shows an example of how the \cbar\ algorithm automatically sets the detection threshold based on the underlying spectral continuum and records the peaks above the threshold itself. For the project, we set a 3.5$\sigma$ detection threshold assuming a number of trials equal to the number of FFT frequencies in each PSD.

In principle, the number of trial periods that one should consider is the total number of Fourier frequencies in all the FFTs carried out in the whole project. However, there are two main reasons why we did not abide by this precept. One is that the total number of searched sources and Fourier frequencies will be unknown until the end of the project. The second is that there is a good number of sources which have been observed more than once with \cxo\ and/or \xmm\ (or with other missions). Thus, we preferred to pre-screen the candidate signals based solely on the statistical properties of each individual time series so to leave open the possibility to check later for the recurrence of the same signal within the \cbar\ project or for confirmations from data from other missions (mainly \xmm, \swift, \asca, and \rst). {\it A posteriori}, this approach proved to be rather efficient, with about 10 signals confirmed by further \cxo\ pointings carried out during the 15 years interval of our project, and about 15 signals confirmed by data from other missions.  

The process described above was implemented in a pipeline which, using tools of the \cxo\ Interactive Analysis of Observations (\textsc{ciao}; \citealt{fruscione06}) software package, retrieves the relevant ACIS data, detects the sources, extracts the event list files for each object, applies barycentric correction to the event times, checks for the presence of significant (source intrinsic) signals, and, when signals are found, infers their main properties.  In particular, we used the \textsc{ciao} releases from 4.4 to 4.8 (as the project started in 2012 June); the task \textsc{axbary} was used for the barycentrization, the tasks \textsc{ardlib, mkinstmap, mkexpmap}, and \textsc{makepsfmap}, for constructing the exposure and point-spread function maps, and \textsc{wavdetect} for the source detection. For the latter, we considered five wavelet detection scales (1, 2, 4, 8, and 16 pixels), and the default value for the source pixel threshold (\textsc{sigthresh} parameter set to $10^{-6}$). We notice that the choice of the latter parameters essentially does not impact the results of the project, since we are interested only in relatively `bright' point-like sources, with at least 50 counts. This minimum number of counts is dictated by the intrinsic capability of the FFT to detect a 100\% modulated signal above a 3$\sigma$ threshold level. {\rc Finally, each source event list was extracted from the {\rcb the 3$\sigma$-distribution}  region { \core determined} by \textsc{wavdetect}.} Besides our codes and \textsc{ciao}, for the timing analysis we also made use of a number of \textsc{heasoft} tools \citep{blackburn95}, such as \textsc{powspec, lcurve, efsearch}, and \textsc{efold}. 

Particular attention was paid to the presence of instrumental signals originated by the spacecraft dithering (see also Section\,\ref{spurious}).  A first check is automatically performed by the \cbar\ pipeline every time a significant peak is recorded. The procedure is based on the \textsc{ciao} task \textsc{dither\_region}\footnote{See http://cxc.harvard.edu/ciao/ahelp/dither\_region.html.} and checks whether an artificial signal is present in the time series due to the variation of the effective area within the source extraction region. The right panel of Fig.\,\ref{dps} shows an example of a FFT of the time variation of the effective area (as for the FFT in the left panel, the solid blue line marks the 3.5$\sigma$ detection threshold). In the example, the highest peak corresponds to the well-known ACIS dithering period of 707~s followed by several higher harmonics. The comparison between the spectra, as in the example given in the two panels of Fig.\,\ref{dps}, offers to us an objective tool to reject or validate a candidate signal.  {\rc Additionally, we also inspected the PSD { \core computed from time series of photons accumulated in background regions around or close to} each new pulsator (the size and shape of the background region depend from case to case) to check for the presence of peaks consistent with the detected signal.}

As of 2015 December 31, we extracted about 430\,000 time series from sources with more than 10 counts (after background subtraction); $\sim$190\,000 of them have more than 50 counts and their PSDs were searched for significant peaks. At the time of writing, the total number of searched Fourier frequencies was about $4.3\times10^9$. After a detailed screening, we obtained a final sample of 41 (42) new X--ray pulsators (signals), which are listed in Table\,\ref{maintable}. For completeness we also list in Table\,\ref{data} the identification number of all the data sets we used for the analysis. None of the recorded spurious signals falls at the frequency of those periods. 
The above reported numbers make \cbar\ the largest search for signals carried out so far in soft X-rays. Moreover, since the \cxo\ mission is still operative, \cbar\ has to be considered a living project: every 1--3 months, the pipeline  will be applied to the new public data sets. Based on the results obtained so far, we expect to find about three new pulsators every year.  

It is worth emphasizing that the FFT is particularly efficient when dealing with relatively high-statistics time series and/or strong signals. Alternative timing tools, such as  periodograms based on the $Z_n^2$ test (where $n$ stands for the number of the assumed harmonics of the signal) are better suited for the search of faint signals, or strong signals in low-statistics time series,  and/or non sinusoidal signals \citep{buccheri83}. However, any periodogram algorithm approach suffers from the known problem related to the variable period resolution over the searched period interval. In this respect, the  $Z_n^2$ is better suited when the search is carried out over a relatively narrow period interval (about one order of magnitude).  A $Z_n^2$-based algorithm able to deal with at least four order of magnitudes in periods with a variable period resolution will be applied in the future to the whole sample of  430\,000 \cxo\ time series. 

\section{\xmm\ archival data}\label{xmm}
{\rc For each pulsator, when available, we also retrieved, reduced and analysed the relevant \xmm\ archival data of the European Photon Imaging Camera (EPIC). Raw data were reprocessed by means of the Science Analysis Software (\textsc{sas}, version between 12.0 and 14.0). Data 
were filtered { \core for episodes of flaring particle background} and the { \core times affected by these events were} excluded from the analysis. Source photons were extracted from circles with radius of { \core 40\arcsec or less, depending on the presence of nearby sources and/or background issues and/or the distance from the CCD edges}.
EPIC background even lists were extracted from source-free regions, the sizes and shapes of which were  dictated by nearby sources or background { \core matters}. Photon arrival times were converted to the Solar system barycenter using the task \textsc{barycen} using the source coordinates as inferred from the
\cxo\ pointings. The ancillary response files and the spectral redistribution matrices for the spectral analysis
were generated with \textsc{arfgen} and \textsc{rmfgen}, respectively. 
}

\section{Catalogue}\label{catalogue}
The main properties of the newly identified \cxo\ ACIS pulsators, together with their signals, are listed in Table\,\ref{maintable}. 
The columns report the following quantities/values: (1) The name of the source, where the prefix CXO is used for sources listed in 
the \cxo\ Source Catalog\footnote{See http://cxc.harvard.edu/csc/ for more details.} version 1.1 (CSC; \citealt{evans10}), while CXOU 
for yet uncatalogued sources; (2,3) Right ascension and declination (J2000); (4) The Galactic latitude; (5) The period of the detected signal; 
(6) a flag, which summarizes the statistical robustness of the detection (see below for the detailed explanation); (7) The flux of the source or, 
for variable sources, the flux interval within which the pulsations have been detected;  (8) Quantity of Chandra and XMM-Newton observations; 
(9) Any further relevant information. In some particular cases, we added a section to clarify the possible classification or report on noteworthy facts.     
The source name and coordinates (columns 1--3) are those in the CSC for catalogued objects (CXO prefix), and the outcomes of the 
\textsc{wavdetect} task for the uncatalogued objects (CXOU prefix).

The periods and their uncertainties in column (5) were inferred in different ways, using case-by-case the technique deemed more 
appropriate for the data sets available for a source. Depending on the number of observations, their length and count statistics, 
the periods were determined by either a phase-fitting analysis, a folding analysis, a $Z_n^2$ test, or a fit with one or more 
sinusoidal functions. In many cases, due to poor statistics or low number of sampled cycles, we are not able to test if the correct 
periods are the ones detected or twice these values (as sometimes occurs in CVs). The pulsed fraction is defined as:
\mbox{$PF\equiv(F_{\rm{max}}-F_{\rm{min}})/(F_{\rm{max}}+F_{\rm{min}})$}, where $F_{\rm{max}}$ and $F_{\rm{min}}$ are the observed 
count rates at the peak and at the minimum of the pulse\footnote{{\rc See http://www.physics.mcgill.ca/$\sim$aarchiba/pulsed-flux-CASCA-07.pdf 
for a comparison among different definitions of pulsed fraction.}}. 

{ Three different classes of robustness were defined and one has been assigned to each signal (column 6)}. The flags are defined as follow. Three asterisks (***) identify an extremely robust signal, which has been detected in more than one data set. Two asterisks (**) indicate a signal detected in only one dataset (the only one available for the given source), but at a high confidence level (above 5$\sigma$). Finally, one asterisk (*) is assigned at those signals detected either at a relatively low confidence level (in between 3.5$\sigma$ and 5$\sigma$), or only in one dataset out of two or more observations. Although the latter group of (*) signals 
might be considered less reliable, we note that a large fraction, about 30\%, of the signals first flagged (*) or (**) were later upgraded to the flag (***) after the analysis of further archival or follow-up X-ray observations. This group may also include transient sources.  

The values in column 7 are observed fluxes in the 0.5--10~keV, as measured from background-subtracted source spectra, which were created
with the \textsc{ciao}'s \textsc{specextract} task. Energy channels were grouped so to have at least 15 counts in each new energy bin. 
All bins consistent with zero counts after background subtraction were removed before fitting the data. For simplicity and uniformity, 
we always used a simple absorbed power-law model, which in most cases provided an acceptable fit to the data.

The last column of Table\,\ref{maintable} (9) includes references and information deemed interesting for the nature of the source and/or 
the signal. Among others, are the Galactic latitude; possible associations with star clusters, or galaxies, or coincidence with sources
detected from other missions; relevant properties inferred from the 
\begin{landscape}
\begin{table}
\caption{CATS catalogue. An updated online version of the Table is maintained at http://www.oa-roma.inaf.it/HEAG/catsatbar/\,.} \label{maintable}
\begin{tabular}{@{}llllllccl}
\hline
Name & R.A. & Dec. & b& Period$^a$ & Flag$^b$ & Flux$^c$  & XMM/Chandra$^{d}$  & Comments$^{e,f}$ \\
&(hh mm ss.s) & (\degr \,\,\, \arcmin \,\,\, \arcsec)&(\degr)& (s)& &(\flux) &&  \\
\hline
CXO\phantom{U}\,J002415.9--720436 & 00 24 15.9 & --72 04 36.4 & --44.9& 8\,649(1) & *** & 0.5--0.8 &C(16)& E, 47\,Tuc \\
CXOU\,J004814.1--731003 & 00 48 14.1 &  --73 10 03.7 &--44.0& 50.669(1) & *** & 4--45 &C(1)/X(2)& T, HMXB, SMC (SXP\,25.5, SXP\,51,
XMMU\,J004814.1--731003) \\
CXOU\,J005048.0--731817 & 00 50 48.0& --73 18 18.2&--43.8& 292.784(5) & *** & 5--22&C(6)& T, SMC, HMXB (RX\,J0058.2--7231, 
XTE\,J0051--727), O (e. l.), [1]\\  
CXOU\,J005440.5--374320 & 00 54 40.5 &  --37 43 20.2 &--79.4& 21\,180(485) & * & 0.9 &C(1)& T, NGC 300{\rcb, perhaps a foreground CV} \\
CXOU\,J005758.4--722229 & 00 57 58.4  &  --72 22 29.5 &--44.7& 7.918\,07(5)& *** & 4--20 &C(2)/X(1)& T, HMXB, SMC, O (e.l.), [2] \\
CXO\phantom{U}\,J021950.4+570518 &  02 19 50.5 &	+57 05 18.2  &--3.7& 4\,782(5) & *** & 0.3--0.6 &C(4)/X(1)& Pst, h\,Persei (NGC\,869), O (e. l.)\\
CXOU\,J055930.5--523833 & 05 59 30.5 &  --52 38 33.5   &--28.9& 21\,668(200) & * & 0.2 &C(3)&  CV?\\
CXOU\,J063805.3--801854 &   06 38 05.3 & --80 18 54.1    &--27.3& 13\,151(330)  & **  & 4  &C(1)& CV? \\ 
CXOU\,J091539.0--495312 & 09 15 39.0  & --49 53 12.9 &--49.9& 55.97(1) & * & 1.2 &C(1)& transient \\ 
CXOU\,J111133.7--603723 &  11 11 33.7 &  --60 37 23.0  &-60.6&  9\,766.8(5)  & *** & 1.2 &C(7)/X(1)& O (3 possible counterparts) \\ 
CXOU\,J112347.4--591834  &  11 23 47.4 &  --59 18 34.2  &--59.3& 1\,525.92(6)  & ***   & 2.8--5.5  &C(7)/X(2)& near SNR G292.0+1.8 \\
CXO\phantom{U}\,J123030.3+413853 &   12 30 30.3 & +41 38 53.1    &+74.9& 23\,148(396)  & ***  & 0.2--1  &C(5)/X(2)& NGC\,4490, BH--WR candidate [3]\\
CXOU\,J123823.4--682207 & 12 38 23.4  & --68 22 07.0 &--5.5& 5\,065(40) & * & 2.1 &C(1)& \\
CXOU\,J141332.9--651756  &  14 13 32.9         &  --65 17 56.5   &--3.8&  6\,377(4)  &  ***   & 0.4--0.6   &C(2)/X(1)& CV (Polar?) [4] \\
CXO\phantom{U}\,J141430.1--651621   & 14 14 30.1  & --65 16 23.3 &--3.8& 6\,120(2)& *** & 1.1--1.4 &C(4)/X(1)& IP [4] \\
  & & && 64\,200(500) & ** && & Same source as above. Long period, likely of orbital origin\\
CXOU\,J153539.8--503501 &    15 35 39.8 &   --50 35 01.4   &--4.2&   12\,334(567)  &  **   &  9 &C(1)&  NGC\,5946  \\
CXO\phantom{U}\,J161437.8--222723 & 16 14 37.9 & --22 27 23.7 &+20.2& 5\,671(137)  & ** & 0.5 &C(1)&  T, CV?, 100\% PF\\
CXOU\,J163855.1--470145 &  16 38 55.1 &  --47 01 45.8 &--0.1&  5\,684(40)  & ***  &  0.9 &C(2)/X(1)& NS?, Norma Arm?, 100\% PF    \\
CXO\phantom{U}\,J170113.3+640757	& 17 01 13.3 & +64 07 58.5 &+36.2& 5\,674(1) & ***  &  0.3 &C(9)/X(1)&  100\% PF,  O (e.l.)\\ 
CXO\phantom{U}\,J170214.7--295933 &  17 02 14.7 & --29 59 33.6 &+7.2& 11\,467(500) & *** & 0.2 &C(6)& Pst, 100\% PF\\ 
CXO\phantom{U}\,J170227.3--484507	& 17 02 27.4 & --48 45 07.1 &--4.3& 3\,080(40) & *** & 0.7--1.4 &C(2)/X(2)& Pst \\
CXO\phantom{U}\,J171004.6--321205	 & 17 10 04.6 &  --32 12 05.5 &+4.5& 4\,990(15)& *** & 0.9--2.8 &C(2)&  $\sim$100\% PF\\
CXOU\,J173037.7--212633 & 17 30 37.7 &  --21 26 33.0 &+4.6& 5\,059.40(7) & *** & 0.3 &C(9)&  near Kepler SNR\\
CXOU\,J173113.7--212552 & 17 31 13.7 & --21 25 52.1 &+6.7& 15\,532(64) & *  & 11 &C(2)&  Pst, O (e. l.)  \\
CXO\phantom{U}\,J173359.0--220614 & 17 33 59.1 & --22 06 14.1 &+5.8& 4\,745(23) & * & 0.4 &C(1)& O \\
CXOU\,J174042.3--534029 & 17 40 42.3  & --53 40 28.9 &--12.0& 21\,134(33) & *** & 5.7 &C(5)&  in  NGC\,6397 \\ 
CXO\phantom{U}\,J174245.1--293455	& 17 42 45.1 & --29 34 55.4 &+0.2& 12\,220(1164) & * & 0.5 &C(2)& Galactic bulge  \\
{\rc CXO\phantom{U}\,J174638.0--285325} &{\rc  17 46 38.0} &{\rc  --28 53 25.9} &{\rc  --0.2} &{\rc  21\,887(538)} &{\rc  ***} &{\rc  2.0} &{\rc C(4)/X(2})&{\rc  Galactic bulge, CV, O (e.l.)} \\
CXOU\,J174811.0--244930 &  17 48 10.1 & --24 49 03.0   &+1.6&  5\,017(19)  &  ***  & 0.4   &C(2)& Terzan\,5, $\sim$100\% PF  \\
CXOU\,J180839.8--274131 & 18 08 39.8  & --27 41 31.6 &--9.9& 854(14)  & *   & 27  &C(1)&  O (e. l.; OGLE source, 0.28\,d)  \\
CXO\phantom{U}\,J180900.0--435039 & 18 09 00.1 & --43 50 40.0 &--11.4& 5\,842(115)  & *  & 1.6  &C(1)&  NGC\,6541 \\
CXOU\,J181516.4--270851 &  18 15 16.4 &  --27 08 51.6 &--4.8& 472(1)  & **   &  4.5  &C(2)& O ?  \\
CXOU\,J181924.1--170607 &  18 19 24.1 &   --17 06 07.2 &--0.9& 407.8(1)  & ***  & 8.5--46  &C(1)/X(3)&  V, O (e.l.),  [5,6]\\
CXO\phantom{U}\,J184441.7--030549	& 18 44 41.7 & --03 05 49.4 &+0.1& 6\,366(236) & ** & 0.4-11 &C(3)/X(3)& V   \\
CXOU\,J185415.8--085641 & 18 54 15.8  & --08 56 41.2 &--4.7& 5\,790(208) & *  & 0.6  &C(1)&    \\
CXOU\,J191043.7+091629 & 19 10 43.7  & +09 16 29.2 &--0.02& 36\,204(109)& ** & 4--24 &C(2)/X(1)& V, HMXB, { AX J1910.7+0917, INTEGRAL},
$K=11.8$\\   
CXO\phantom{U}\,J193437.8+302524	& 19 34 37.8 & +30 25 24.4 &+5.0& 5\,906(200) & *** & 1.8 &C(1)/X(1)& O (e.l.)  \\
CXOU\,J204734.8+300105 & 20 47 34.8  & +30 01 05.2 &--8.4& 6\,097(82) & *** & 9--21 &C(1)/X(1)& E, Polar ?,  O (e.l.)  \\
CXOU\,J215447.8+623155 & 21 54 47.8 & +62 31 55.0 &+6.3& 9\,933(10) & *** & 1.6  &C(4)&  O (e.l.) \\ 
CXOU\,J215544.5+380116 & 21 55 44.5 &   +38 01 16.3 &--13.0& 14\,090(43) & ** & 0.3 &C(2)& O (e.l.)  \\ 
CXOU\,J225355.1+624336 & 22 53 55.1 & +62 43 36.8 &+2.9& 46.673\,66(4) & *** & 20--50 &C(5)/X(1)&  HMXB, O (e. l.), [8] \\
\hline
\end{tabular}
\end{table}
\end{landscape}

\begin{landscape}
\begin{table}
\contcaption{} \label{maintable2}
\begin{tabular}{@{}llllllccl}
\hline
Name & R.A. & Dec. & b& Period$^a$ & Flag$^b$ & Flux$^c$  & XMM/Chandra$^{d}$ & Comments$^{e,f}$ \\
&(hh mm ss.s) & (\degr \,\,\, \arcmin \,\,\, \arcsec)&(\degr)& (s)& &(\flux) &&  \\
\hline
\multicolumn{9}{c}{Sources with signals published by other groups while the project was on-going and/or with rough estimate of the periods}\\
\hline
CXO\phantom{U}\,J165334.4--414423 &  16 53 34.4 & --41 44 23.8  &+1.3& 5\,820(20)  & ***   & 0.2   &C(1)/X(6)& CV,  [9] \\
CXO\phantom{U}\,J174728.1--321443 & 17 47 28.1 & --32 14 43.9 &--2.1& 4\,831(2) & *** & 0.4--9.5 &C(3)/X(1)& V, [7]\\
CXO\phantom{U}\,J182531.4--144036 & 18 25 31.5 & --14 40 36.5 &--1.1& 781(1) & *** & 14 &C(2)/X(1)& [7], CV ? \\
CXO\phantom{U}\,J191404.2+095258 & 19 14 04.2 & +09 52 58.4 &--0.5& 5\,937(219) & * & 280 &C(1)/X(1)&  HMXB, IGR\,J19140+0951,
$P{\rm orb}=13.55$\,d, $K{\rm s}=8.7$, [10] \\
\hline
\end{tabular}
\begin{list}{}{}
\item[$^{a}$] The uncertainty is given at a 90\% confidence level.
\item[$^{b}$] The number of stars indicates detection confidence as follows: *** Multiple detection of the same signal (either with \cxo\ or with other missions); ** Single high confidence ($>$10$\sigma$) detection; * Single detection (above 3.5$\sigma$). See also Section\,\ref{catalogue}.
\item[$^{c}$] Observed 0.5--10~keV flux in units of $10^{-13}$~\flux\  for a power-law fit. A range of values indicates different values measured in the \cxo\ observations used for the timing analysis. 
\item[$^{d}$] {\rc Number of useful \cxo\ and \xmm\  pointings used to evaluate the signal/source parameters. 
We counted only those observations where the signal was detected or those with no detection but 3$\sigma$ pulsed fraction upper limits smaller than that of the modulation were inferred. We also counted observations where the source 
was not detected at a comparable or lower flux level with respect to the original detection level or detected at a significantly different flux level.}
\item[$^{e}$] E: eclipsing; T: transient: the source was not detected in some observations with \cxo\ or other missions; Pst: persistent: the flux did not vary by more than a factor of 5 in the observations we considered; V: variable: we measured flux variations larger than a factor of 5; O: optical counterpart or association. Question marks denote tentative identifications or associations.
\item[$^{f}$] References: [1] \citet{eisrc13}; [2] \citet{bartlett16}; [3] \citet{eism13}; [4] \citet{eimm15};  [5] \citet{nichelli11b}; [6] \citet{farrell15}; [7] \citet{muno08}; [8] \citet{eis13}; [9] \citet{lin14};  [10] \citet{sidoli16}.
\end{list}
\end{table}
\end{landscape}
\noindent X-ray data (transient, eclipsing, persistent, or variable sources) 
or optical data (both from catalogues or follow-up observations).    

For a subsample of 10 new X-ray pulsators, we obtained X-ray follow-up observations with \cxo, \swift, or \xmm, seven of which were 
already carried out. Similarly, in the latest four years, we performed optical spectroscopic follow-up for about half the sources. 
In those cases where a good optical counterpart candidate was found, based on the positional coincidence and the detection of emission 
lines (e.l.) in the spectrum, we included the piece of information in column 9 as `O(e.l.)'. More details  and results from X-ray and 
optical follow-up observations are already published for some sources (the references are provided in column 9 and in the paragraphs 
on individual sources) {\rc and for others will be included in a forthcoming paper (Israel et al. in preparation)}.  

In column 9 we used the following acronyms: NS, neutron star; BH, black hole; WR, Wolf--Rayet star; SNR, supernova remnant; SMC, Small Magellanic Cloud. 
Most sources are expected to be CVs or HMXBs; in general we favour a CV identification when the source is far from the Galactic plane 
(high or low Galactic latitude $b$) and/or the luminosity appears to be low ($<$$10^{32}$~\lum) and/or there are possible associations 
with globular clusters or faint optical counterparts; we favour an HMXB nature when a period variation ($\dot{P}$) is measured and/or 
there is a possible bright optical counterpart. The pulse profiles of detected pulsators are shown in Fig.\,\ref{efold1} and \ref{efold4} together with their inferred periods and pulsed fractions.

An on-line version of the \cbar\ catalogue is maintained at http://www.oa-roma.inaf.it/HEAG/catsatbar/\,. It will include new sources 
detected in the future and updated information.

\subsection{Individual sources}

The signals reported for CXO\,J165334.4--414423, CXOU\,J181924.1--170607, CXO\,J182531.4--144036 and CXO\,J174728.1--321443 were independently discovered in other projects using \xmm\ or \cxo\ data \citep{lin14,farrell15,muno08}. We listed them even so, because either we discovered them before their publication (for the first two cases; \citealt{lin14,farrell15}) or because our analysis led us to different 
conclusions (in the other instances; \citealt{farrell15,muno08}).  

\subsubsection{CXO\,J002415.9--720436}
This source is possibly part of the globular cluster 47\,Tuc, which was observed with \cxo\ many times (e.g. \citealt{grindlay01,edmonds03,egh03,heinke05}; \citealt*{hge05}). The timing and spectral parameters in Table\,\ref{maintable} were inferred from observations 953, 955, and 2735--8. Although the signal was originally detected at about 4\,325~s, the study of the pulse shape revealed that the correct period is about 8\,650~s. In fact, when folded at the latter period, the  pulse profile is asymmetric and shows a total eclipse lasting for about 8 minutes (Fig.\,\ref{efold1}). \citet{edmonds03} proposed an X-ray period of 6.287~h, which is not confirmed by our analysis. The length of the period and the presence of a total eclipse strongly favour a CV interpretation, as it was proposed by \citet{edmonds03}.
\subsubsection{CXOU\,J004814.1--731003} The power spectrum shows one highly-significant signal at 51~s plus its harmonic at 25.5~s. Moreover, the \cxo\ position is consistent with that of XMMU\,J004814.1--731003, a 25.5-s Be/XRB pulsar \citep{haberl08}. Our findings strongly suggest that SXP\,25.5 \citep{lamb02}, SXP\,51 \citep{galache08}, XMMU\,J004814.1--731003, and CXOU\,J004814.1-731003 are all the same source, and its true period is $\sim$51~s.
\subsubsection{CXOU\,J005048.0--731817} 
CXOU\,J005048.0--731817 is a transient HMXB pulsar in the SMC. In 2010 April--May, this source underwent a long-lived ($\sim$20~d) outburst, during which the luminosity increased by about two orders of magnitude (up to $\sim$ $1.2\times10^{36}$~\lum). The event was fortuitously sampled by 7 \cxo\ observations and the source was serendipitously detected in its quiescent state, at a luminosity of $\sim$ $10^{34}$~\lum, in other 15 \cxo\ pointings from 2002 October to 2010 June. The fact that, despite all this, the source went unnoticed in the \cxo\ data for years, even in campaigns focussed on the study of HMXBs in the SMC, testifies to the importance of systematic searches for periodic signals.

Detailed analysis of this transient pulsar was reported in \citet{eisrc13}. Here we just make two remarks. Firstly, we note that   CXOU\,J0050 is a different object from the SMC pulsar RX\,J0058.2--7231 \citep*{hep08}, which has a similar period ($\sim$291~s) but is located about 1$^\circ$ away. However, it is unclear whether  CXOU\,J0050 or  RX\,J0058.2--7231, as proposed by \citet{hep08}, is responsible for the RXTE detection of the 293~s signal tentatively catalogued as  SXP\,293 (XTE\,J0051--727; \citealt{corbet04}), whose uncertain position is consistent with both objects \citep{galache08}. Secondly, the analysis of archival  spectra of the optical counterpart of CXOU\,J0050 taken in September 2007 with the 3.6-m ESO telescope, allowed us to refine the spectral classification, which resulted to be a O9--B0 (more likely an O9) IV--V class spectral-type star. 
\subsubsection{CXOU\,J005758.4--722229} Detailed analysis of this source will appear in \cite{bartlett16}.
\subsubsection{CXO\,J021950.4+570518} The presence of a faint ($V>20$) counterpart in the X-ray error circle and a luminosity of (2$-4)\times10^{30}$~\lum\ (assuming a distance to NGC\,869 of 2.4~kpc; \citealt{currie09}) suggest a CV.
\subsubsection{CXOU\,J063805.3--801854} A bright star ($B=16.3$, $R=15.7$; USNO-B1.0 0096-0020652) is consistent with the X-ray error circle. 
\subsubsection{CXO\,J123030.3+413853}
CXO\,J123030.3+413853 is located in the low-metallicity spiral galaxy NGC\,4490, which is interacting with the smaller irregular NGC\,4485. In \citet{eism13}, we interpreted its strong modulation at $\sim$6.4~h ($\sim$90\% pulsed fraction, confirmed also by \xmm\ data) as an orbital period and, considering the morphology of the folded light curve and the maximum peak X-ray 
\begin{figure*}
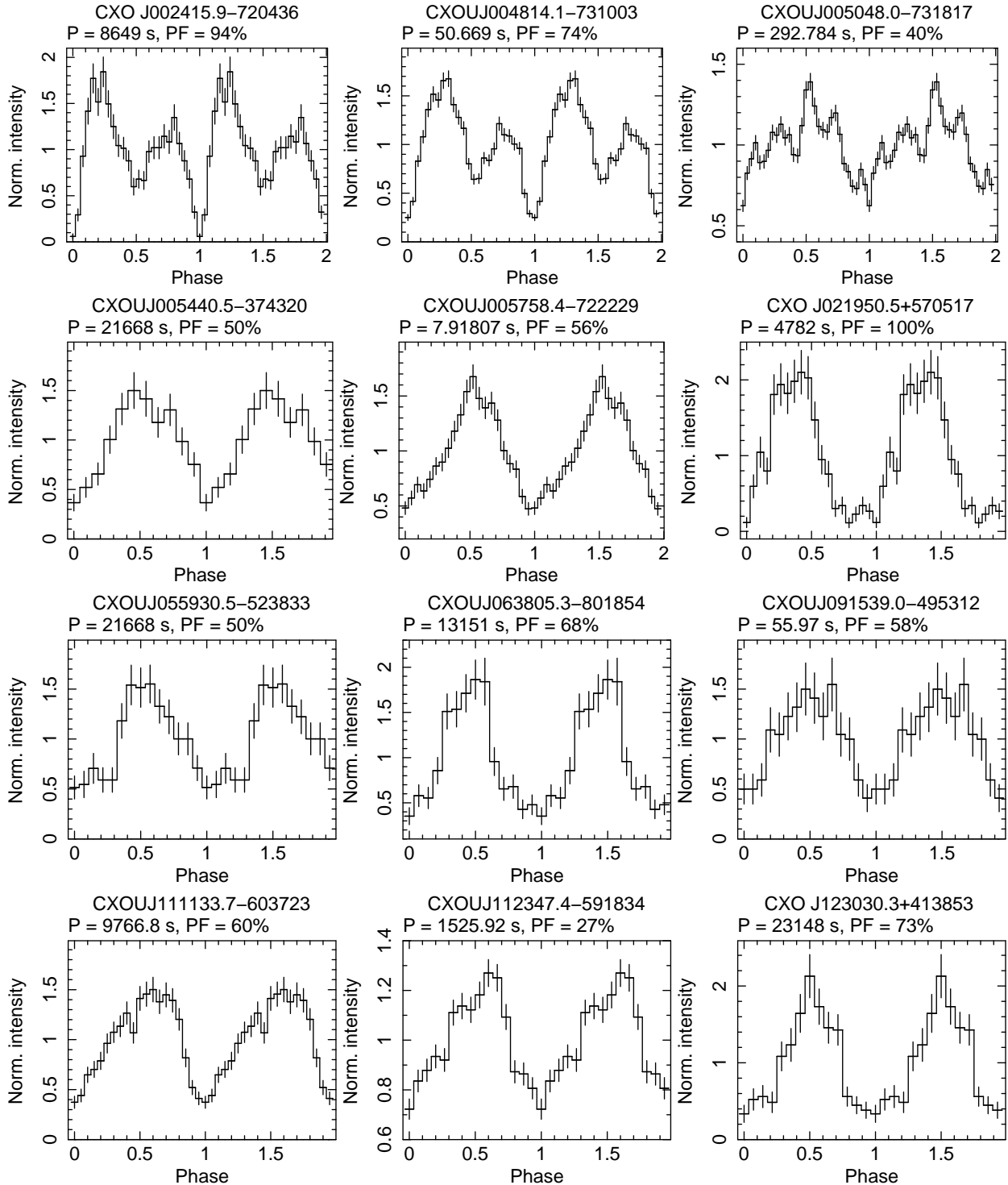

\centering
\resizebox{0.93 \hsize}{!}{\includegraphics[angle=-90]{C0024E.ps}\hspace{2mm}\includegraphics[angle=-90]{C0048E.ps}\hspace{2mm}\includegraphics[angle=-90]{C0050E.ps}}\vspace{1mm}
\resizebox{0.93 \hsize}{!}{\includegraphics[angle=-90]{C0054E.ps}\hspace{2mm}\includegraphics[angle=-90]{C0057E.ps}\hspace{2mm}\includegraphics[angle=-90]{C0219E.ps}}\vspace{1mm}
\resizebox{0.93 \hsize}{!}{\includegraphics[angle=-90]{C0559E.ps}\hspace{2mm}\includegraphics[angle=-90]{C0638E.ps}\hspace{2mm}\includegraphics[angle=-90]{C0915E.ps}}\vspace{1mm}
\resizebox{0.93 \hsize}{!}{\includegraphics[angle=-90]{C1111E.ps}\hspace{2mm}\includegraphics[angle=-90]{C1123E.ps}\hspace{2mm}\includegraphics[angle=-90]{C1230E.ps}}
\caption{\label{efold1} CATS pulse profiles. The source, as well as the period used to fold the data and the pulsed fraction, are indicated in each panel.}
\end{figure*}
\begin{figure*}
\centering
\resizebox{0.96 \hsize}{!}{\includegraphics[angle=-90]{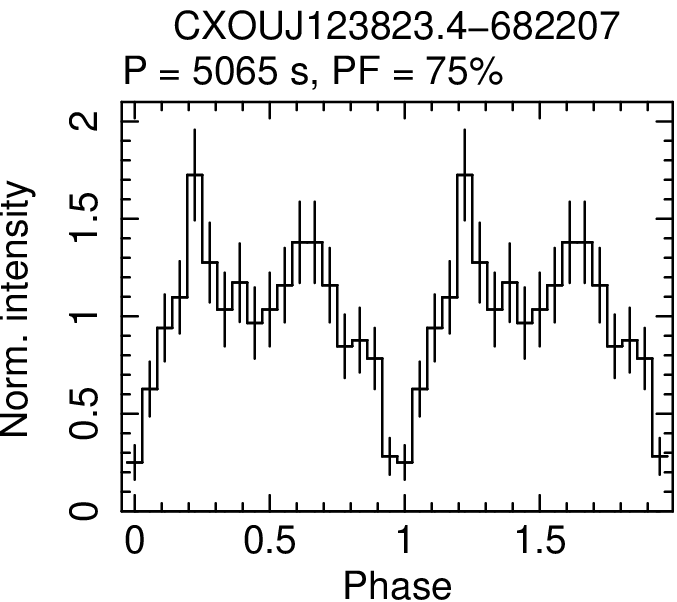}\hspace{2mm}\includegraphics[angle=-90]{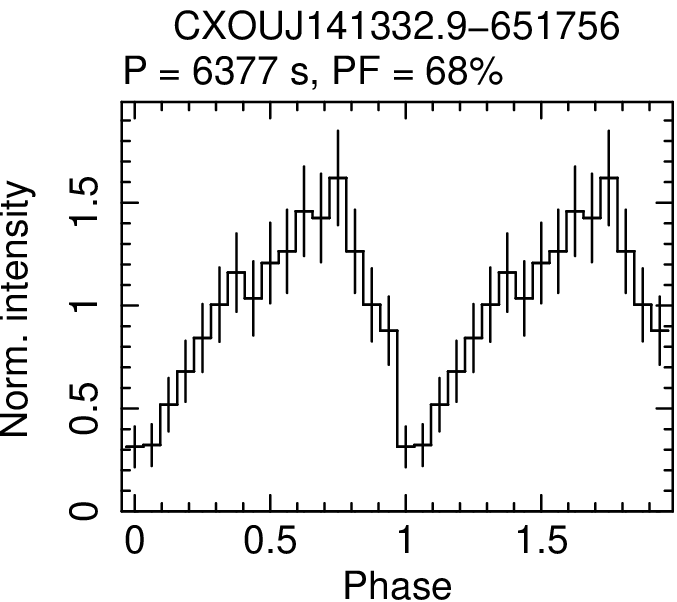}\hspace{2mm}\includegraphics[angle=-90]{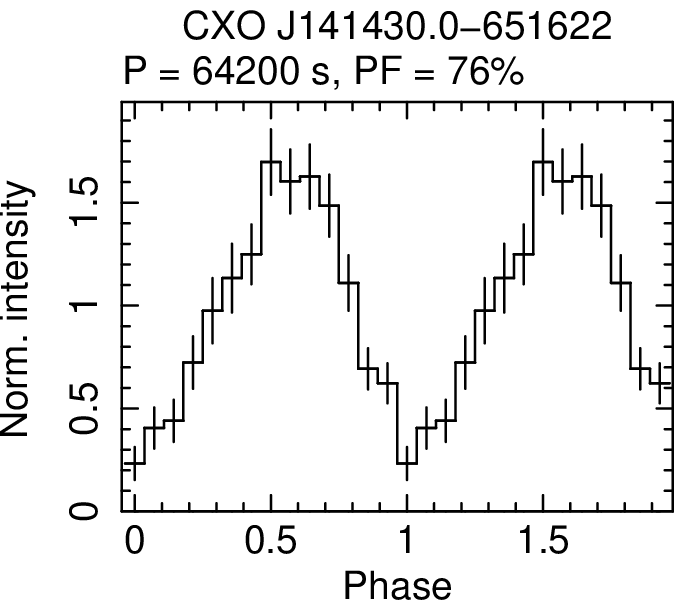}}\vspace{1mm}
\resizebox{0.96 \hsize}{!}{\includegraphics[angle=-90]{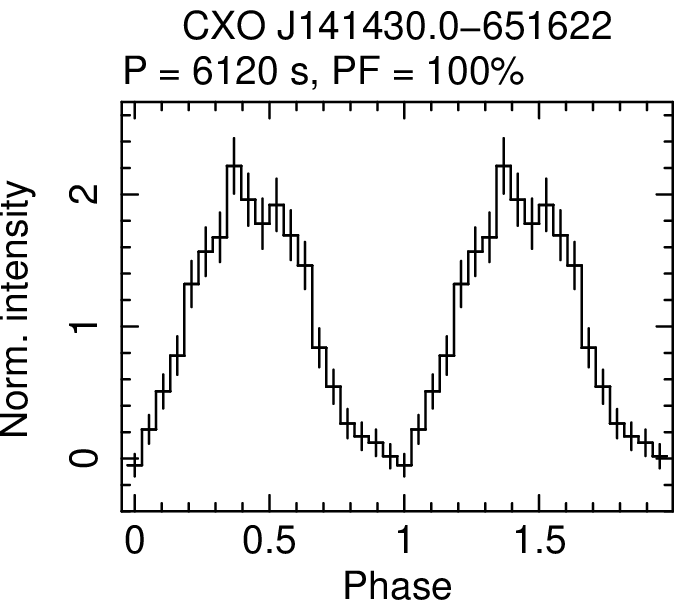}\hspace{2mm}\includegraphics[angle=-90]{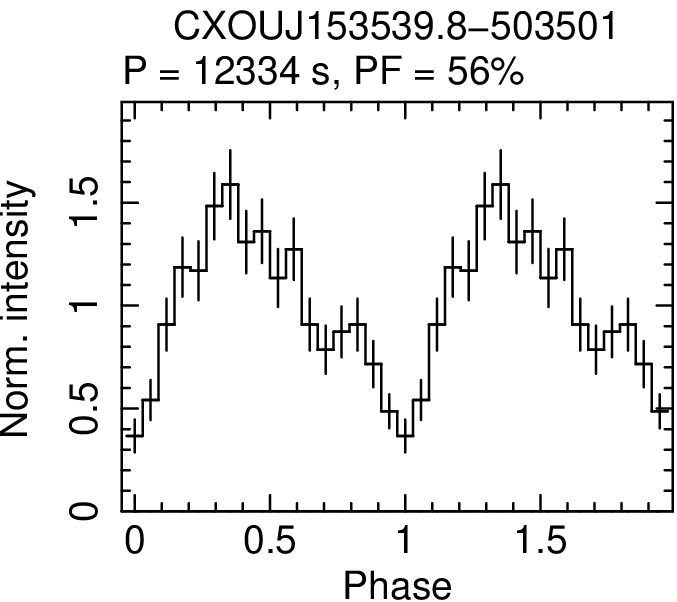}\hspace{2mm}\includegraphics[angle=-90]{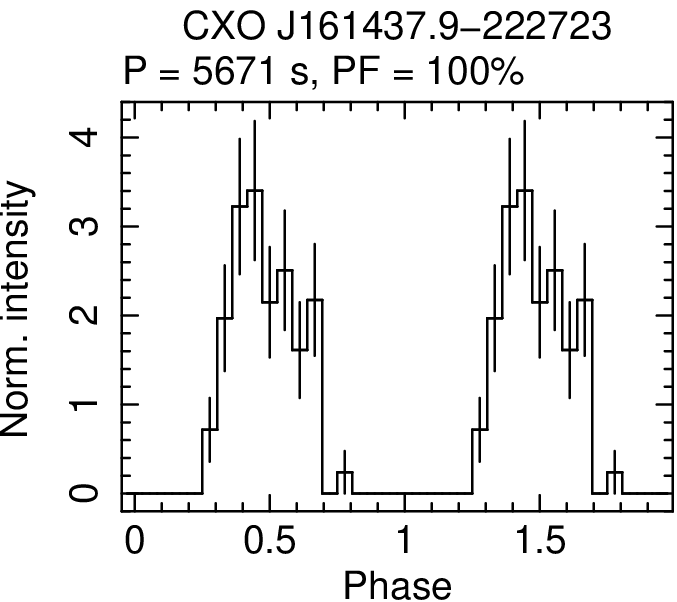}}\vspace{1mm}
\resizebox{0.96 \hsize}{!}{\includegraphics[angle=-90]{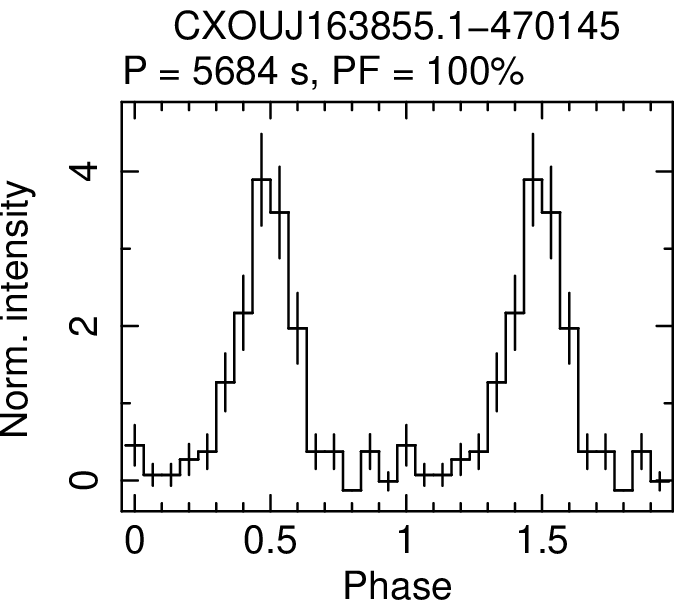}\hspace{2mm}\includegraphics[angle=-90]{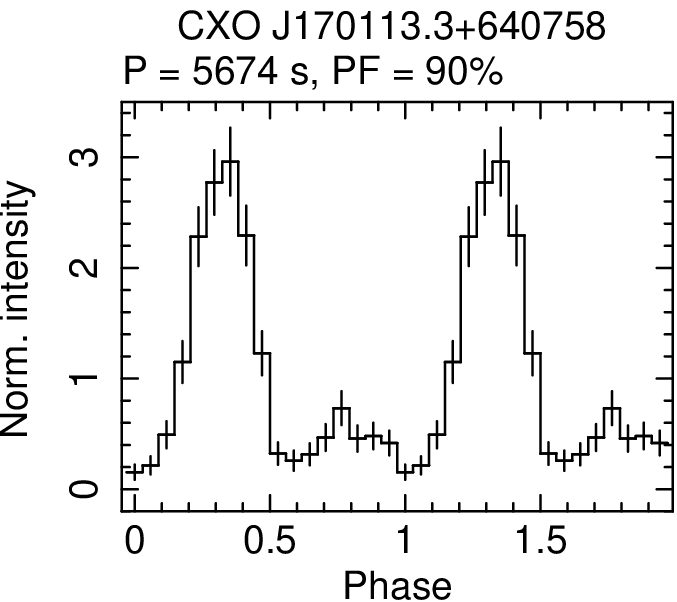}\hspace{2mm}\includegraphics[angle=-90]{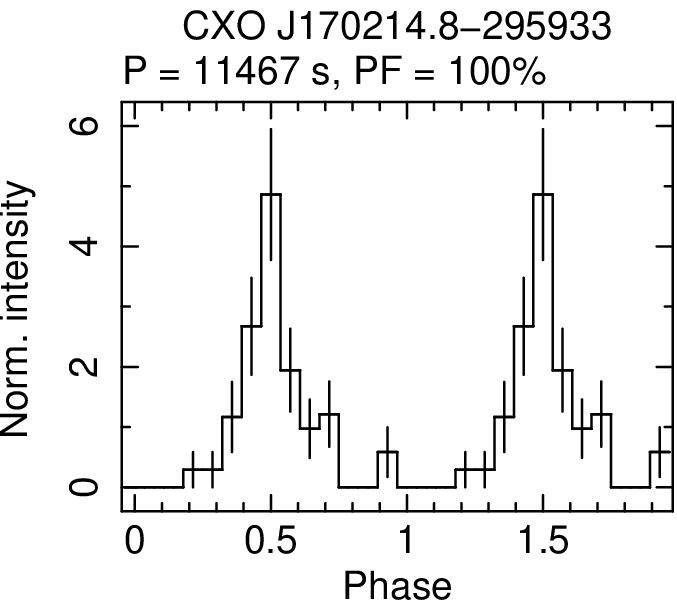}}\vspace{1mm}
\resizebox{0.96 \hsize}{!}{\includegraphics[angle=-90]{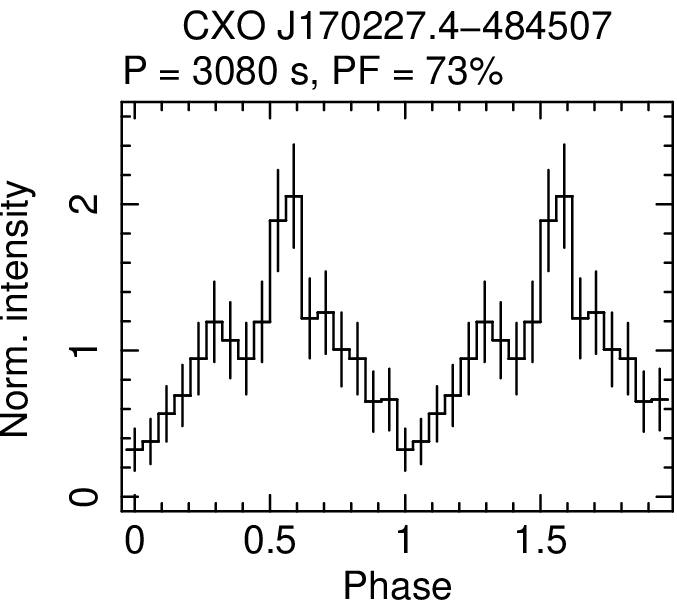}\hspace{2mm}\includegraphics[angle=-90]{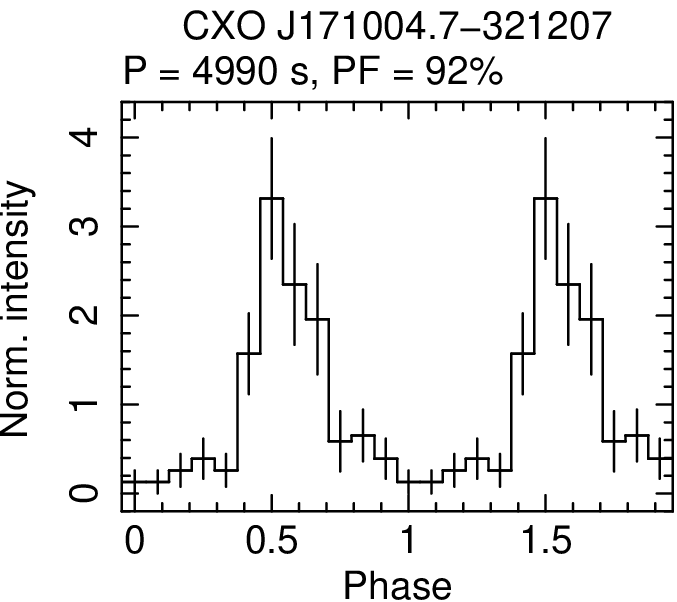}\hspace{2mm}\includegraphics[angle=-90]{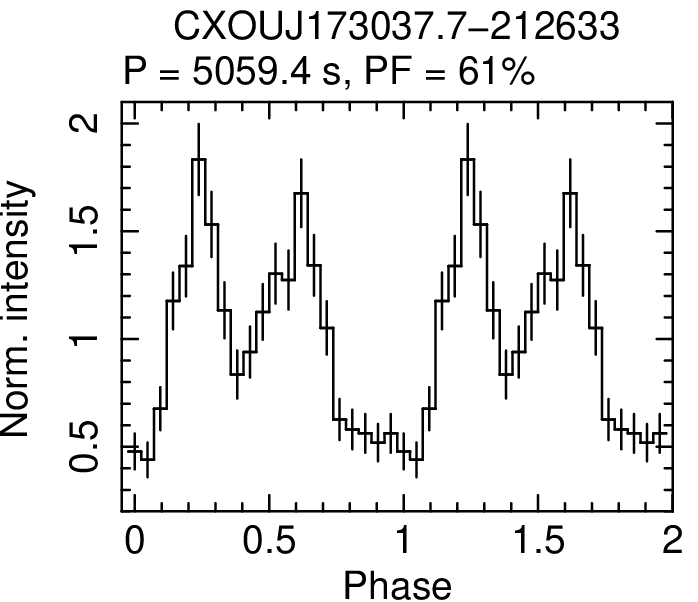}}
\contcaption{\label{efold2}  }
\end{figure*}
\begin{figure*}
\centering
\resizebox{0.96 \hsize}{!}{\includegraphics[angle=-90]{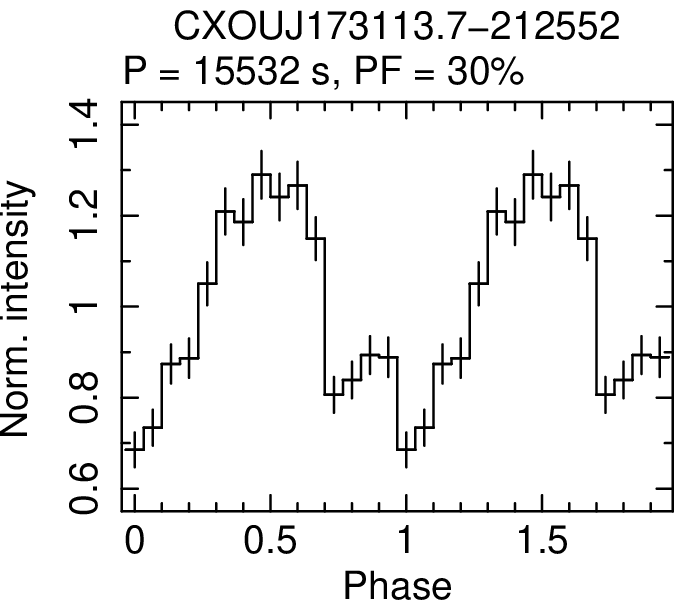}\hspace{2mm}\includegraphics[angle=-90]{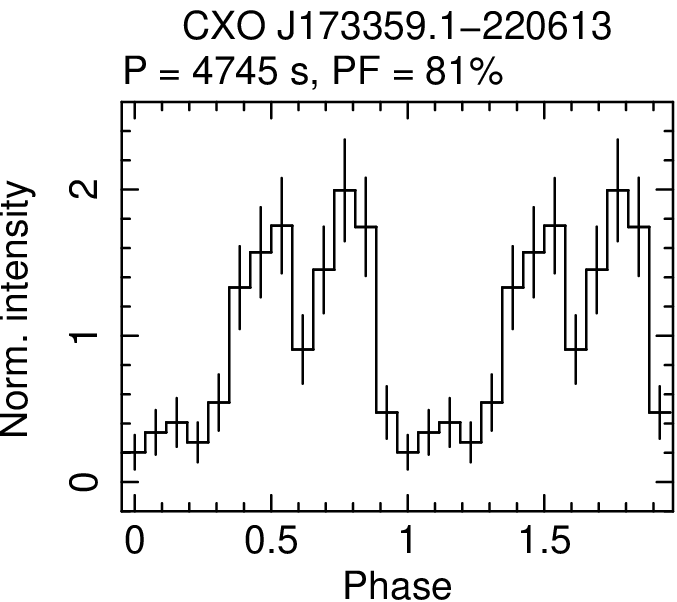}\hspace{2mm}\includegraphics[angle=-90]{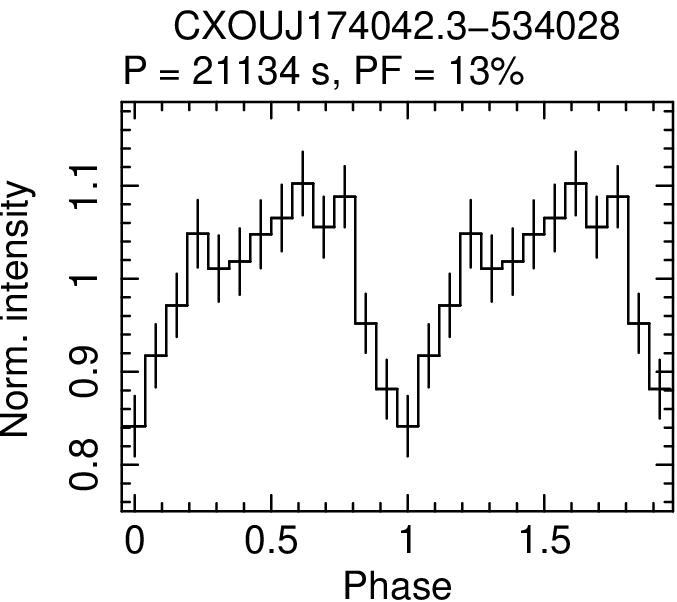}}\vspace{1mm}
\resizebox{0.96 \hsize}{!}{\includegraphics[angle=-90]{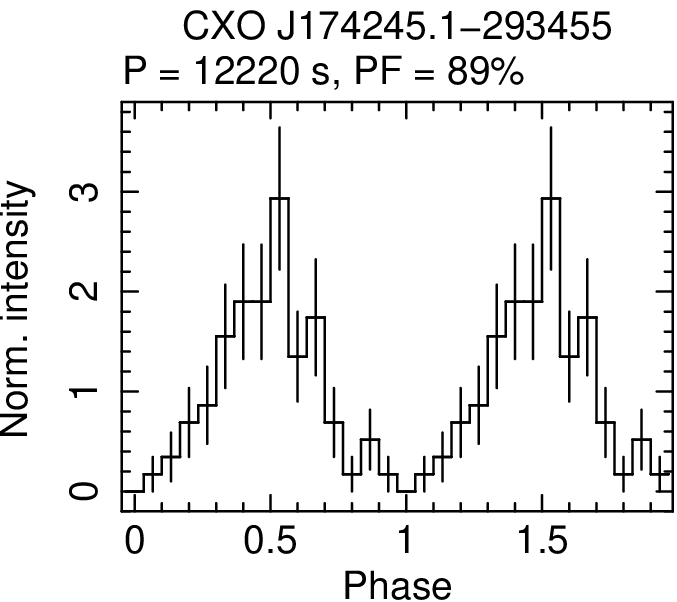}\hspace{2mm}\includegraphics[angle=-90]{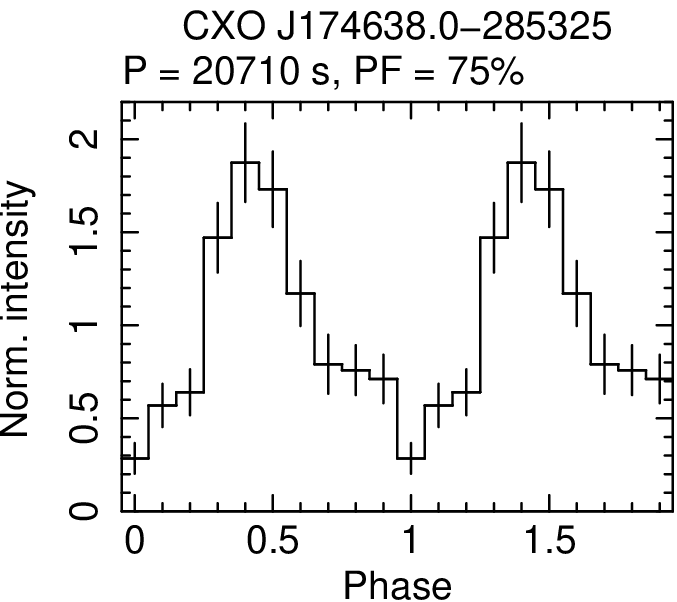}\hspace{2mm}\includegraphics[angle=-90]{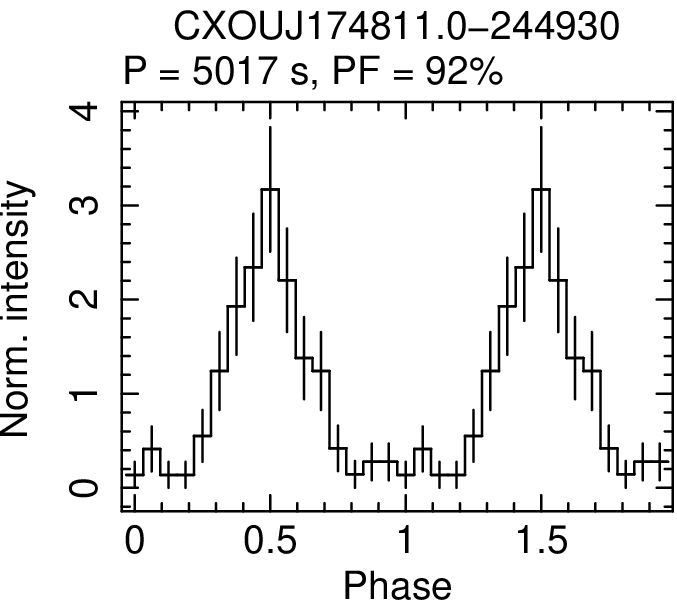}}\vspace{1mm}
\resizebox{0.96 \hsize}{!}{\includegraphics[angle=-90]{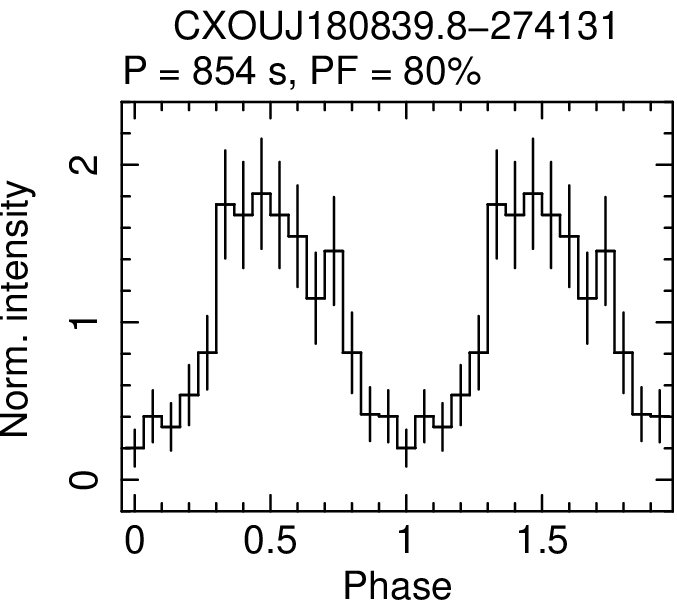}\hspace{2mm}\includegraphics[angle=-90]{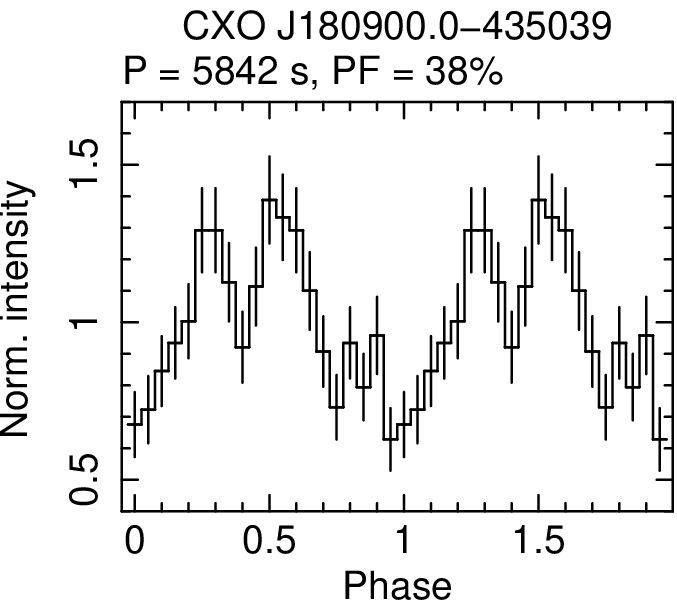}\hspace{2mm}\includegraphics[angle=-90]{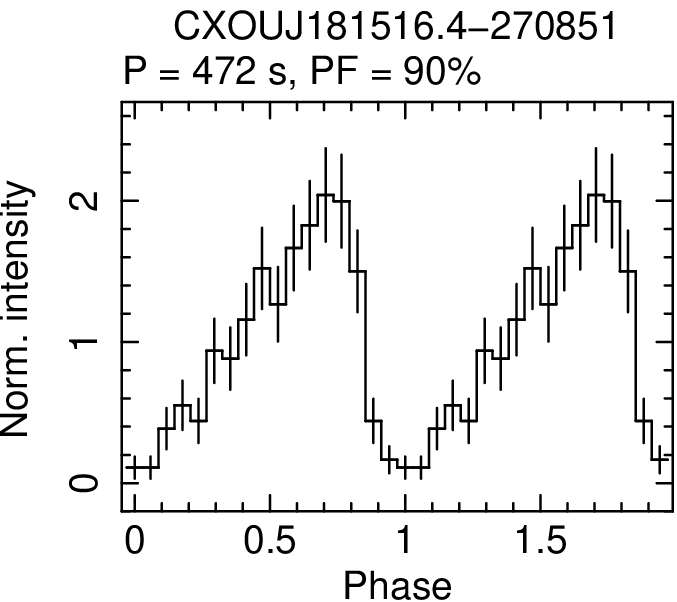}}\vspace{1mm}
\resizebox{0.96 \hsize}{!}{\includegraphics[angle=-90]{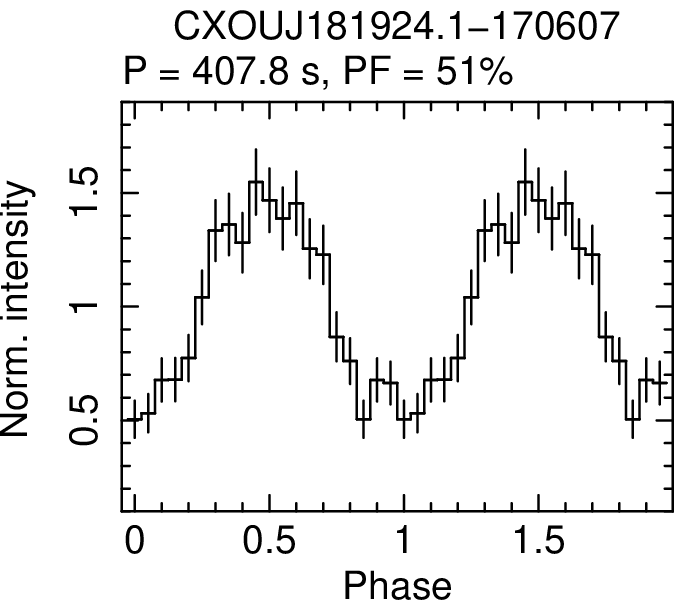}\hspace{2mm}\includegraphics[angle=-90]{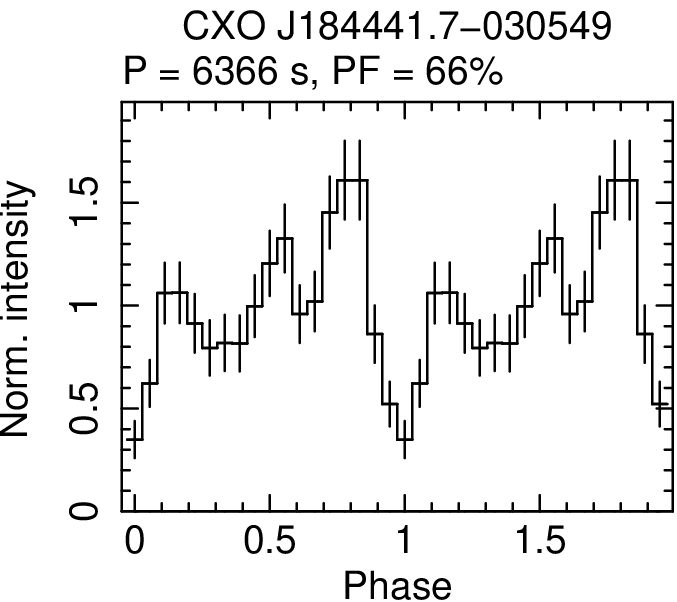}\hspace{2mm}\includegraphics[angle=-90]{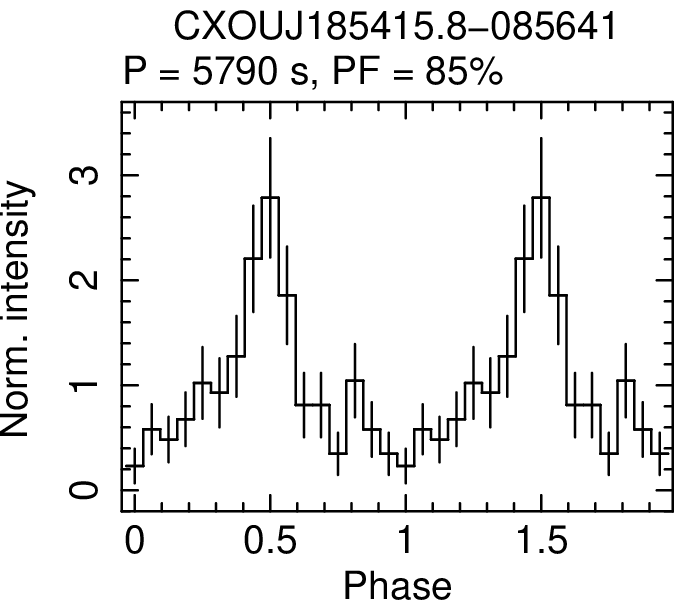}}
\contcaption{\label{efold3}  }
\end{figure*}
\begin{figure*}
\centering
\resizebox{0.96 \hsize}{!}{\includegraphics[angle=-90]{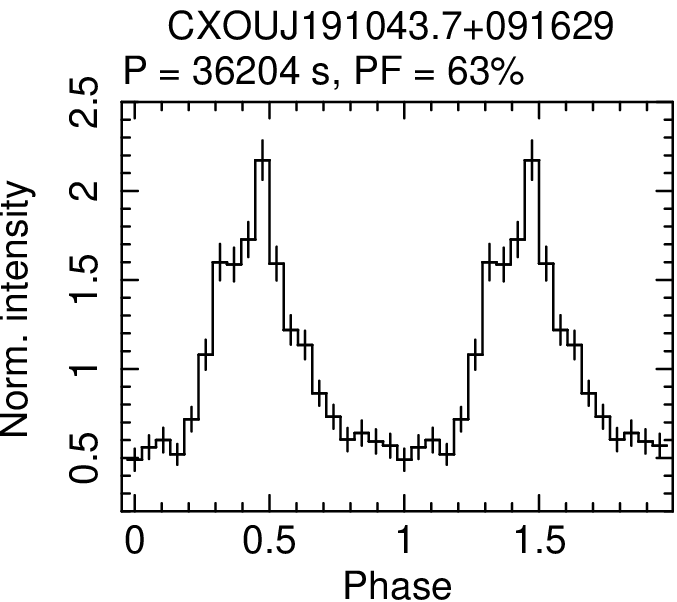}\hspace{2mm}\includegraphics[angle=-90]{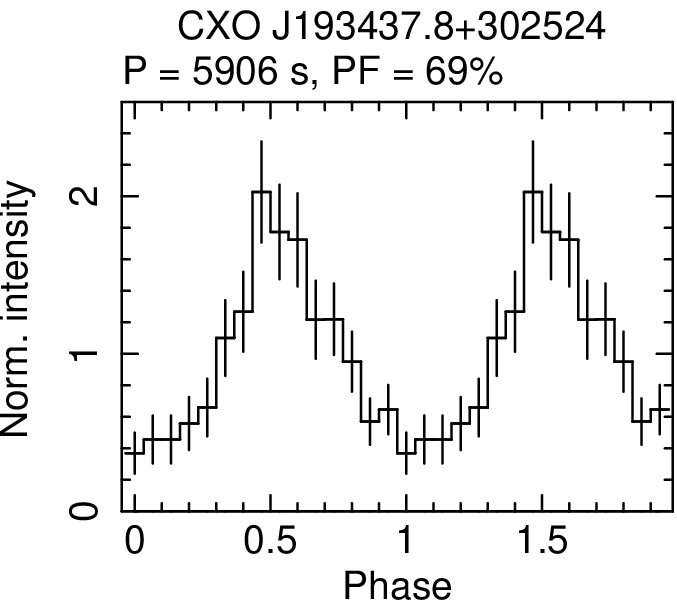}\hspace{2mm}\includegraphics[angle=-90]{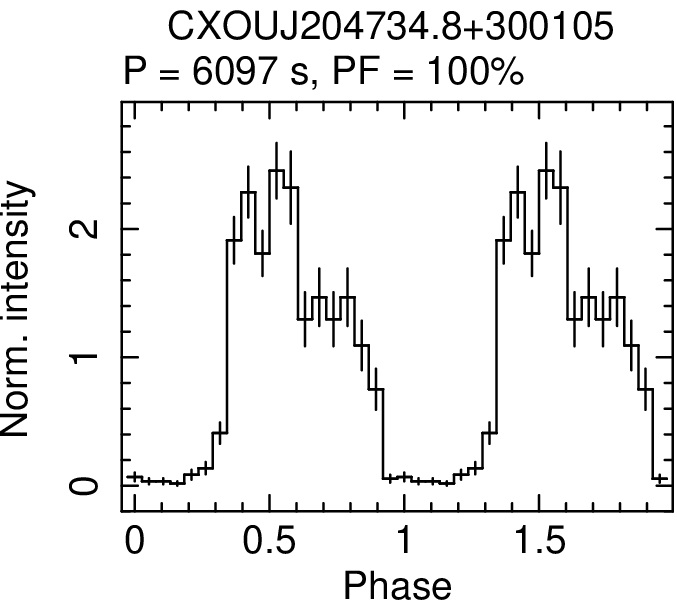}}\vspace{1mm}
\resizebox{0.96 \hsize}{!}{\includegraphics[angle=-90]{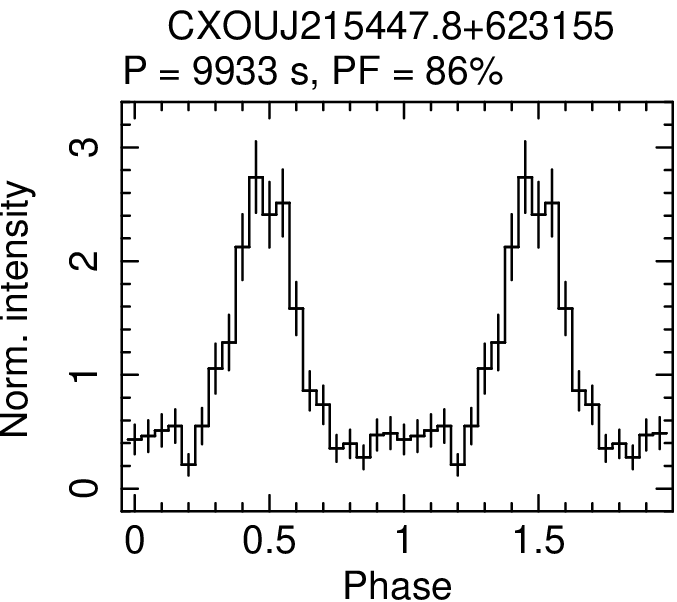}\hspace{2mm}\includegraphics[angle=-90]{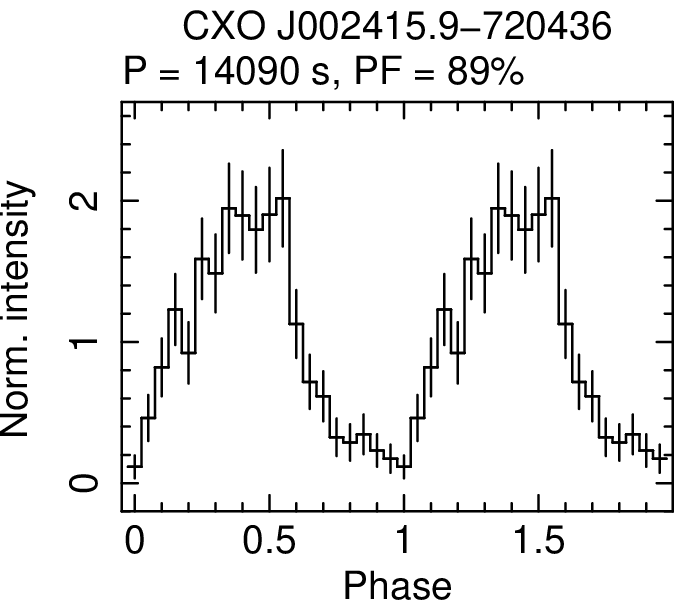}\hspace{2mm}\includegraphics[angle=-90]{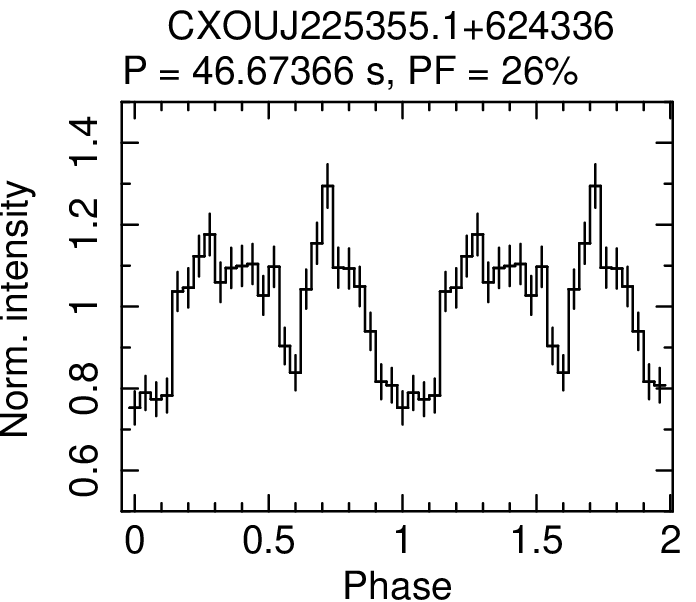}}
\contcaption{\label{efold3b}  }
\end{figure*}
\noindent  luminosity of $\approx$ $2\times10^{39}$~\lum\ (for $d=8$~Mpc), we proposed the source as a BH HMXB with a Wolf--Rayet (WR) star donor.

Besides their rarity and evolutionary interest, these systems are particularly interesting because they are precursors of NS--BH and BH--BH binaries, which may merge emitting gravitational waves and creating a more massive BH \citep{abbott16}. In \citet{eimm15}, we derived from the current sample of BH--WR candidates (7 objects, in four of which the WR counterpart has been securely identified) an upper limit to the detection rate 
of stellar BH--BH mergers with Advanced LIGO and Virgo of $\sim$16~yr$^{-1}$ 
\subsubsection{CXOU\,J141332.9--651756  and CXO\,J141430.1--651621} Periodic signals from these sources were discovered in a long exposure of the Circinus galaxy (ESO\,97--G13).
CXO\,J141430.1--651621 is more than 2~arcmin out of the border of the Circinus galaxy and therefore an association is ruled out. A spin period of 1.7~h and an orbital period of 17.8~h pin down the source a CV of the IP type. Its X-ray spectrum can be modelled by a power law with photon index $\Gamma\simeq1.4$ and the flux of $\approx$$1\times10^{-13}$~\flux\ shows $\approx$50~per~cent variability on time-scales of weeks--years. The typical luminosity of IPs and the nondetection of the optical counterpart suggest a distance larger than $\sim$5~kpc \citep{eimm15}. 
Albeit at $\sim$3.5~arcmin from the nucleus, CXOU\,J141332.9--651756 appears inside the Circinus galaxy. However, the absorption column, which is much smaller than the total Galactic density in that direction, argues against an extragalactic source. Indeed, the probability of a foreground Galactic X-ray source is substantial ($\approx$10~per~cent, as estimated from the cumulative Galactic X-ray source density versus flux distribution from the \cxo\ Multi-wavelength Plane survey by \citealt{vandenberg12}). The modulation period is 1.8~h and the emission (flux of $\approx$$5\times10^{-14}$~\flux, with $\approx$50~per~cent variations on weekly/yearly scales) can be described by a power law with $\Gamma\simeq0.9$. CXOU\,J141332.9--651756 is probably a (Galactic) magnetic cataclysmic variable, probably of the polar type \citep{eimm15}. Assuming that the companion is a M5V star (or similar), the nondetection of its optical counterpart implies $d \ga 0.7$~kpc; therefore, if the system is within a few kpc, its luminosity is in the normal range for polars.
\subsubsection{CXOU\,J153539.8--503501} This source is within the tidal radius of the globular cluster NGC\,5946 \citep{davidge95}. A possible faint counterpart ($V>20$) is consistent with its \cxo\ error circle.
\subsubsection{CXOU\,J163855.1--470145} Archival \xmm\ data confirm the modulation at a period of 5\,827(19)\,s, implying a spin-down rate of $8(2)\times10^{-7}$~s~s$^{-1}$; or 25(8)~s yr$^{-1}$. This suggests that the pulsations reflect the rotation of a NS. Assuming a distance of 10~kpc for the Norma Arm, the luminosity is $\sim$$10^{33}$~\lum.
\subsubsection{CXO\,J170113.3+640757} Optical spectroscopic follow-up observations  led to the identification of an optical counterpart showing H and  He emission lines. 
\subsubsection{CXOU\,J173113.7--212552} The period was discovered in the \cxo\ obs.~6714, but not in obs.~6716, few days apart. An \xmm\ pointing detected the source, but is too short to confirm the signal (less than 2 cycles are covered). Optical spectroscopic follow-up observations carried out at NOT led to the identification of an optical counterpart showing H and He emission lines.
\subsubsection{CXO\,J174638.0--285325} {\rc \citet{koenig08} reported on the optical spectroscopy of the likely counterpart of this source as part of the study of the \cxo\ Multiwavelength Plane Survey, ChaMPlane. Based on broad H$\alpha$ emission line in the spectrum of the { \core $R=20.2$} magnitudes counterpart and a X-ray to optical flux ratio of { \core 0.07,} \citet{koenig08} classified this source as a candidate CV. Our discovery of  a period of $\sim$6.1 hr { \core supports a CV nature for} the new pulsator.}
\subsubsection{CXO\,J174728.1--321443} \citet{muno08} proposed a modulation around 5\,000~s using the same data sets (4566--4567, 2004 March). In an \xmm\ follow-up observation carried out in 2014 August, we measured a period of 4\,941(52)~s which is, within uncertainties, consistent with the \cxo\ value.  \xmm\  caught the source at a much lower flux of $\sim$ $4\times10^{-14}$~\flux, a factor of about 25 fainter. 
\begin{figure*}
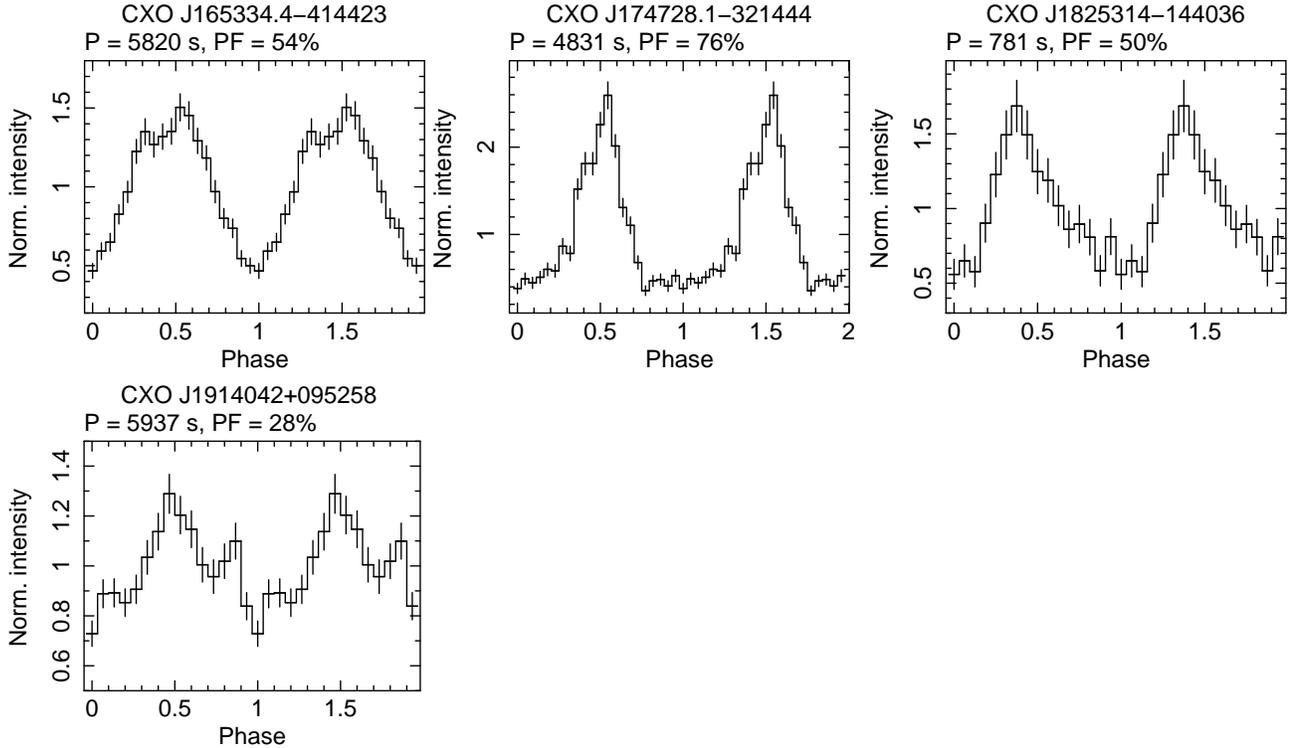

\centering
\resizebox{0.96 \hsize}{!}{\includegraphics[angle=-90]{C1653E.ps}\hspace{2mm}\includegraphics[angle=-90]{C1747E.ps}\hspace{2mm}\includegraphics[angle=-90]{C1825E.ps}}\vspace{1mm}
\resizebox{0.31 \hsize}{!}{\includegraphics[angle=-90]{C1914E.ps}}\hspace{11.5cm}
\caption{\label{efold4}  Same as Figure\,\ref{efold1} but for the latest three sources in Table\,\ref{maintable}}
\end{figure*}
\subsubsection{CXOU\,J174811.0--244930, CXO\,J180900.0--435039} These sources are within the tidal radius of the globular cluster Terzan\,5 \citep{lanzoni10} and NGC\,6541 \citep{harris96}, respectively.
%
%
\subsubsection{CXOU\,J181924.1--170607} First discovered by our group in \swift\ and \xmm\ datasets \citep{nichelli11b}, was later rediscovered  by \citet{farrell15}. Optical spectroscopic follow-up observations carried out at NOT led to the identification of a faint optical counterpart showing H and He emission lines. Furthermore, by using all the \cxo, \xmm\ and \swift\ archival data, which span over more than 8 years, we inferred a 3$\sigma$ upper limit on the first period derivative of $|\dot{P}|< 10^{-11}$\,s\,s$^{-1}$. These findings strongly disfavour the scenario of a slow X-ray pulsar in a HMXB \citep{farrell15}, while strengthens the CV scenario proposed by \citet{nichelli11b}.
\subsubsection{CXO\,J182531.4--144036} {\rc Pulsations at a period of 785(2)\,s and 5\,000\,s were first reported by \citet{muno08}, based on a \xmm\  pointing carried out on 2008 April 10--11, { \core when} the source was detected at a flux of $1.9(1)\times10^{-12}$~\flux (35\% larger than in the \cxo\ data). The 5\,000\,s signal was classified as a likely noise fluctuation. Our detection of  781(s)\,s pulsations in a 2004 July  \cxo\ observation shows that the period did not changed significantly among the two epochs. Additionally, there is evidence for a longer modulation around about 5\,000\,s in the same \cxo\ dataset. We assigned a *** flag to the 781\,s signal. }
\subsubsection{CXO\,J191404.2+095258} 
CXO\,J191404.2+095258 is the soft X-ray counterpart of IGR~J19140+0951, discovered with \igr\ in 2003 March \citep{hannikainen03}. It was identified as an HMXB with orbital period of 13.552(3)~d \citep{corbet04,wen06} and a B0.5 supergiant companion \citep{intzand04} located at a distance of about 3.6\,kpc \citep{torrejon10}. 
{\rc It is a very variable source \citep{chr04,intzand04}, 
with an X-ray intensity that can span 3 orders of magnitudes \citep{sidoli16}}.
\citet{intzand04} reported on the hint of a possible modulation at $\sim$6.5~ks, that needed confirmation. 
A modulation near that value was indeed found in our project ($\sim$5.9~ks; Table\,\ref{maintable}).
{\rc A 1.46 mHz quasi-periodic oscillation was discovered during a recent XMM-Newton observation \citep{sidoli16}.}
\subsubsection{CXO\,J193437.8+302524, CXOU\,J204734.8+300105, CXOU\,J215447.8+623155, CXOU\,J215544.5+380116}
For all these sources, optical spectroscopic follow-up observations led to the identification of optical counterparts showing H and He emission lines. 
\subsubsection{CXOU\,J225355.1+624336}
We discovered in CXOU\,J225355.1+624336 a modulation at $\sim$47~s in a series of \cxo\ observations carried out in 2009. The modulation was recovered also in \swift\ and \rst\ data, allowing us to infer an average increasing rate of the period of $\sim$17~ms per year in 16 years and, therefore, to decipher the signal as the rotation period of an accreting, spinning-down NS \citep{eis13}.

Follow-up observations at the Nordic Optical Telescope made it possible to classify the NS companion as a B0-1III-Ve (most likely a B1Ve) star at a distance of about 4--5 kpc. The X-ray luminosity of $\sim$ $3\times10^{34}$~\lum, steady within a factor of $\sim$2, suggests that 1RXS\,J225352.8+624354 is a new member of the sub-class of low-luminosity persistent Be/X-ray pulsars similar to X Persei, which have long orbital period ($P_{\mathrm{Orb}}\ga30$~d) and wide and circular orbits ($e<0.2$).
\begin{figure*}
\centering
\resizebox{0.7 \hsize}{!}{\includegraphics[angle=-90]{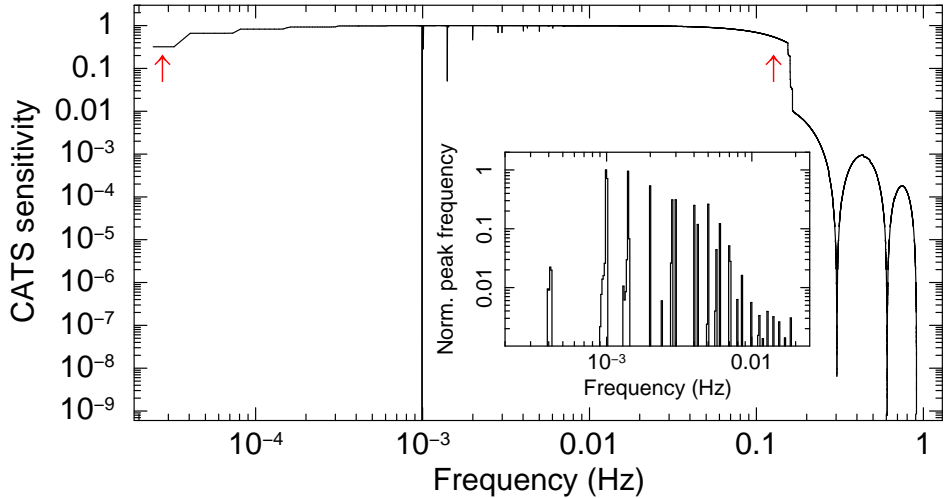}}
\caption{\label{spur}  {\rc Sensitivity map as a function of frequency} for the CATS signal search: drops in the sensitivity are due to the dithering signals {\rc (in the 10$^{-3}$-- 10$^{-2}$\,Hz interval)} and the sampling times ({\rc above 0.1\,Hz;} see Section\,\ref{spurious} for details).  The red arrows mark the longest and shortest period discovered by the CATS pipeline. In the inset we show the more frequent spurious signals detected by the pipeline during the search, the most prominent being at the dithering frequencies plus many harmonics.}
\end{figure*}
\begin{figure*}
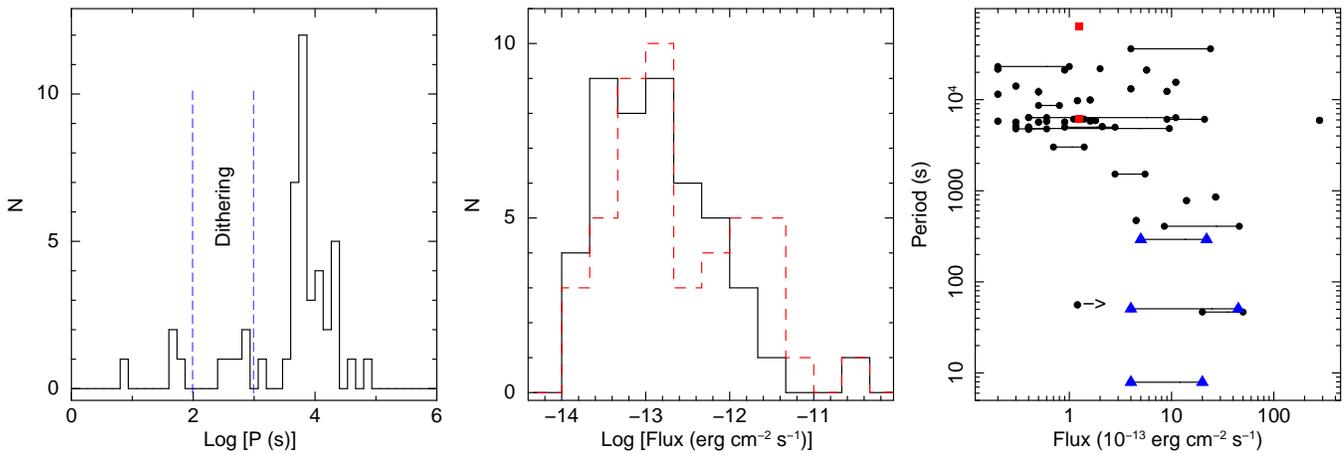

\centering
\resizebox{\hsize}{!}{\includegraphics[angle=-90]{Perdistr2.ps}\hspace{.5cm}\includegraphics[angle=-90]{Fluxdistr.ps}\hspace{.5cm}\includegraphics[angle=-90]{PvsFluxNEW2.ps}}
\caption{\label{catstat}  Main properties of the \cbar\ sample of new X-ray pulsators. From left to the right are shown the period distribution (left panel), the  flux distribution (central panel), and the period versus flux plane (right panel), where  the solid lines mark the objects detected at different flux levels (the right arrow marks a variable object for which we can rely upon only one dataset), the (blue) triangles corresponds to the transient NSs in the Be X-ray binaries of the SMC, and the (red) squares the spin and orbital periods of the new CV/IP in the field of view of the Circinus galaxy. {\rc In the left panel we marked the period interval where the effects of the spacecraft dithering humper the detection of true signals.} In the central panel the solid black and the stepped red lines correspond to the flux distributions obtained by assuming the minimum and maximum fluxes of variable objects, respectively.}
\end{figure*}

\section{Signals of instrumental origin} \label{spurious}
{\rc As a by-product of the { \core project,} we collected and analysed all the information related to  the spurious signals recorded by the detection algorithm.}
Fig.\,\ref{spur} shows the {\rc  sensitivity map as a function of frequency} for the CATS signal search. Boundaries are dictated towards low frequencies by the maximum length of a single pointing (one orbit of about 130\,ks), and at high frequencies by the readout times of the ACIS. For most  observations it is 3.2\,s (0.3125\,Hz), while in  subarray modes it can be from 1/2  to 1/8 of the nominal value; this is reflected by the bumps at high frequency which are due to the corresponding Nyquist frequencies. 
The sharp drops are produced by the ACIS dithering periods (1\,000\,s and 707\,s) and their harmonics extending up to 0.02\,Hz. This is shown in more 
details in the inset, where we plotted all the detected spurious signals. 
We notice the presence of a number of additional signals, which are not likely associated to any of the fundamental dithering frequencies but present in some pointings (such as the signal at 4$\times$10$^{-4}$~Hz). 
The use of the  \textsc{ciao} task \textsc{dither\_region} for each detected signal allowed us to reject these cases (see also Section\,\ref{methodology}).
The red arrows mark the longest and shortest period discovered by the CATS pipeline, namely CXOU\,J005758.4--722229 (7.9\,s) and CXOU\,J191043.7+091629 (36\,204\,s, see Table\,\ref{maintable}).

\section{Summary and conclusions}
In Figure\,\ref{catstat} we present some characteristics of the \cbar\ selected sample. The first panel shows the period distribution. Although some frequency intervals are affected by the sensitivity issues discussed in Section\,\ref{spurious}, a clustering around $\log P \simeq 3.8$ (about 6\,000~s) is evident. This resembles the period distribution of CV peaking at    $\log P \simeq 3.8$ and $\log P \simeq 4.2$ with a gap in between \citep{knigge11}. This agrees with the preliminary classification of the sources in Table\,\ref{maintable}. 

It is worth noticing that all the similar projects carried out in the past and briefly overviewed in Section\,\ref{intro} discovered {\rc only pulsators with periods above $\sim$400\,s, we also detected pulsators as fast as 8\,s. This is partly due to the algorithm employed (mainly LS periodograms and algorithms based on the Z$^{2}$ statistics), partly might be due to the likely larger number of pulsators with intrinsically longer periods, and partly 
to the increasingly incidence of spin modulations at short periods. In the latter { \core case,} their detection depends on the physical conditions of the accreting object (see below).} 

The observed flux distribution (in the 0.5--10\,keV band) is shown in the second panel of Figure\,\ref{catstat}. The relatively high number of new pulsators with low fluxes likely reflects the average flux of the \cxo\ point sources \citep{evans10,wang16}. At any rate, this is a testament of the effectiveness of periodicity search as a tool to infer the nature of faint objects.

The third panel of Figure\,\ref{catstat}, where the periods are plotted against the flux, shows an empty region corresponding to sources with short period and low fluxes. {\rc In fact, though we discovered very faint pulsators below 10$^{-13}$\flux\  with periods above $\sim$3\,000s, we did not detect any pulsator with similar fluxes but shorter periods. This lack is not due (only) to poor statistics given that pulsators with similar 
number of photons but longer periods were reported in Table\,\ref{maintable}.} Since there is not an obvious bias against these detections, a possible interpretation could be in terms of  a magnetic gating mechanism  (e.g. \citealt{campana98,perna06} and references therein). {\rc Whether or not a magnetized compact object accretes on its surface depends from the comparison between to distances:  the corotation radius $r_{\mathrm{cor}}$ and the magnetospheric radius $r_{\mathrm{mag}}$. The former marks the distance at which the balance between the centrifugal forces and the local gravity is satisfied and it is defined as:
\begin{equation}
r_{\mathrm{cor}} = \left(\frac{G M_X P^2}{4 \pi^2}\right)^{1/3} 
 \end{equation}
where $M_X$ and $P$ are the mass and spin period, respectively, of the compact object.
The latter marks the distance where the balance between the magnetic pressure and the ram pressure of the infalling material and can be written as follows:
\begin{equation}
 r_{\mathrm{mag}} = \mu^{4/7} (2GM_X)^{-1/7} \dot{M}^{-2/7}
 \end{equation}
where $\mu$ is the magnetic moment of the compact object and  $\dot{M}$ the rate of infall matter \citep{ghosh77}.
From the relations { \core above,} it is evident that} the higher  the magnetic field of the compact objects, the larger  the infalling mass rate must be to win the centrifugal force of the rotating magnetosphere, to accrete efficiency on the surface of the compact objects and to emit X-rays modulated at the spin period (corresponding to the physical condition that the corotation radius is larger than the magnetospheric one,  $r_{\mathrm{cor}} <r_{\mathrm{mag}}$).

In this scenario, the deficiency of short-period persistent pulsators at low fluxes is not surprising (the only four objects falling in the depleted zone are transient or variable sources). Nonetheless, we also warn the reader that the underabundance of accreting NSs with respect to CVs (with intrinsically longer periods and lower fluxes) might also play an important role in depleting the region corresponding to sources with short period and low fluxes.

In principle, a luminosity--spin period diagram might be an useful tool to estimate $r_{\mathrm{mag}}$ and thus to obtain a lower limit on the  magnetic field of the compact object.
However, in order to draw firm conclusions, other key information are needed. In particular the distances, to convert the fluxes into luminosities, and the orbital or spin nature of the modulations (which is not always straightforward). Also, a larger statistics would be desirable to better circumscribe the empty area. 

Such larger statistics is expected to be achieved by means of similar projects which are currently on-going and aimed at exploiting the \swift\ archive (\swift\ Automatic Timing Survey, aka SATS\,@\,BAR) and the \xmm\ data (Exploring the X-ray Transient and variable Sky, aka EXTraS\footnote{See the project web site at \mbox{www.extras-p7.eu}.}). First results from these projects are reported in \cite{eis14,esposito15} and \cite{esposito16}, respectively.

\section*{Acknowledgments} 
The scientific results reported in this article are based on data obtained from the \cxo\ Data Archive. This research has made use of software provided by the \cxo\ X-ray Center (CXC) in the application packages \textsc{ciao}. {\rc We thank { \core the} anonymous referee for valuable comments}. {\rc  \core GLI acknowledges the partial support from ASI (ASI/INAF contracts I/088/06/0 and AAE DA-044, DA-006 and DA-017).}
PE acknowledges funding in the framework of the NWO Vidi award A.2320.0076 (PI: N.~Rea) { \core and is grateful to Patrizia Caraveo for the support received in an important phase of this work}. LS acknowledges financial support from PRIN-INAF 2014. 

\bibliographystyle{mn2e}
\bibliography{biblio}
\bsp
\clearpage
\onecolumn
\appendix
\section{Archival data}
For sake of completeness we list, for each source in Table\,\ref{data}, the corresponding observation(s) we used for the timing analysis and to infer the fluxes. 

\begin{table}
\centering \caption{Datasets used for the timing analysis reported in this work.} \label{data}
\begin{tabular}{@{}lll}
\hline
Name & \multicolumn{2}{c}{Observation ID} \\
 & \cxo & \xmm \\
\hline
CXO\phantom{U}\,J002415.9--720436 &  78	 ,953	 ,955	 ,956	 ,2735	 ,2736	 ,2737	 ,2738	 ,3384	 ,3385  ,3386, & \\ &	 15747	 ,15748	 ,16527	 ,16529	 ,17420 & \dots \\
CXOU\,J004814.1--731003 & 2945	 ,14674 &0110000101,  0403970301,  0404680101\\
CXOU\,J005048.0--731817  &  2945	 ,3907	 ,7156	 ,8479	 ,11095	 ,11096	 ,11097	 ,11980	 ,11981, & \\  
& 11982	 ,11983	 ,11984	 ,11985	 ,12200  ,12208	 ,12210	 ,12211, &\\
& 12212	 ,12215	 ,12216	 ,12217&  \dots \\
CXOU\,J005440.5--374320  & 16028	 ,16029 & \dots  \\
CXOU\,J005758.4--722229  &  13773	 ,14671	 ,15504& 0700580401\\
CXO\phantom{U}\,J021950.4+570518  &  5407	 ,5408	 ,9912	 ,12021& 0201160201 \\
CXOU\,J055930.5--523833  &  12264	 ,13116	 ,13117& 0604010301\\
CXOU\,J063805.3--801854  & 14925 &  \dots \\
CXOU\,J091539.0--495312  &  14544	 ,16603& \dots  \\
CXOU\,J111133.7--603723  &  2782	 ,12972	 ,14822	 ,14823	 ,14824	 ,16496	 ,16497& 0051550101\\
CXOU\,J112347.4--591834   &  126	 ,6677	 ,6678	 ,6679	 ,6680	 ,8221	 ,8447	& 0110012701,  0400330101,  0743980101\\
CXO\phantom{U}\,J123030.3+413853  &  1579	 ,1579	 ,4725	 ,4725	 ,4726& 0112280201,  0556300101\\
CXOU\,J123823.4--682207  &  11308& \dots  \\
CXOU\,J141332.9--651756   &  12823	 ,12824	&  0111240101\\
CXO\phantom{U}\,J141430.1--651621    &  355	 ,356	 ,12823	 ,12824& 	 0111240101\\
CXOU\,J153539.8--503501  &  9956& \dots \\
CXO\phantom{U}\,J161437.8--222723  &  7509& 0404790101\\
CXOU\,J163855.1--470145  &  12516	 ,12517& 0303100101 \\
CXO\phantom{U}\,J170113.3+640757	 &  547	 ,8032	 ,8033	 ,9756	 ,9757	 ,9758	 ,9759	 ,9760	 ,9767& 0107860301\\
CXO\phantom{U}\,J170214.7--295933  &  3795	 ,5469	 ,6337	 ,8984	 ,13711	 ,14453& \dots \\
CXO\phantom{U}\,J170227.3--484507	 &  4445	 ,4446& 0111360501, 0112900201\\
CXO\phantom{U}\,J171004.6--321205	  &  4549, 16708& \dots \\
CXOU\,J173037.7--212633  &  4650	 ,6714	 ,6715	 ,6716	 ,6717	 ,6718	 ,7366	 ,16004	 ,16614& \dots \\
CXOU\,J173113.7--212552  & 6714	 ,6716 & 0084100101\\
CXO\phantom{U}\,J173359.0--220614  & 4583&  \dots \\
CXOU\,J174042.3--534029  &  79,  2668,  2669, 7460,  7461& \dots \\
CXO\phantom{U}\,J174245.1--293455	 &  2283	 ,7043&  \dots \\
CXO\phantom{U}\,J174638.0--285325 & 945, 7048, 14897, 17236 & 0202670801, 0658600101  \\
CXOU\,J174811.0--244930  &  14625	 ,15615& \dots \\
CXOU\,J180839.8--274131  &  15794& \dots \\
CXO\phantom{U}\,J180900.0--435039  & 3779 & \dots \\
CXOU\,J181516.4--270851  &  16505	 ,16506	& \dots \\
CXOU\,J181924.1--170607  &  12348 &0402470101  0604820101,  0693900101\\
CXO\phantom{U}\,J184441.7--030549	 &  8163	 ,11232	 ,11801& 0046540201,  0602350101,  0602350201 \\
CXOU\,J185415.8--085641  &  14585& \dots \\
CXOU\,J191043.7+091629  & 13440	 ,13441	& 0084100401,  0084100501 \\
CXO\phantom{U}\,J193437.8+302524	 &  587& 0723570401 \\
CXOU\,J204734.8+300105  & 740 & 0082540701 \\
CXOU\,J215447.8+623155  &  8938	 ,10818	 ,10819	 ,10820& \dots \\
CXOU\,J215544.5+380116  &  3967	 ,12879	 ,13218	& \dots \\
CXOU\,J225355.1+624336  &  9919	 ,9920	 ,10810	 ,10811	 ,10812& 0743980301\\
\hline
CXO\phantom{U}\,J165334.4--414423  &  6291& 0109490101,  0109490201,  0109490301, \\
                                                        &          &  0109490401,  0109490501,  0109490601\\
CXO\phantom{U}\,J174728.1--321443  &  4566	 ,4567	 ,13580	& 0743980401 \\
CXO\phantom{U}\,J182531.4--144036  &  4600	 ,5341& 0505530101\\
CXO\phantom{U}\,J191404.2+095258  &  4590 & 0761690301\\
\hline
\end{tabular}
\end{table}
\vfill

\label{lastpage}

\end{document}